\documentclass[journal]{IEEEtran}

\usepackage[T1]{fontenc}
\usepackage[utf8]{inputenc}
\usepackage{cite}
\usepackage{amsmath, mathtools}
\usepackage{amsfonts,amsthm,bm}
\usepackage{amssymb}
\usepackage{comment}
\usepackage{graphicx}
\usepackage{standalone}
\usepackage{dsfont}
\usepackage{mathtools}
\usepackage{color, colortbl}
\usepackage{soul}
\usepackage[normalem]{ulem}
\usepackage[acronym,shortcuts]{glossaries}
\usepackage{arydshln}
\usepackage{bbm}
\usepackage{svg} 
\usepackage{algorithm,algorithmic}
\usepackage{adjustbox}
\usepackage[inline]{enumitem}
\usepackage{multirow}
\usepackage{tabularx}
\usepackage{booktabs}

\usepackage{comment}
\usepackage{tikz}
\usepackage{pgfplots}
\pgfplotsset{compat=newest}
\pgfplotsset{plot coordinates/math parser=false}
\newlength\fheight
\newlength\fwidth
\usetikzlibrary{plotmarks,patterns,decorations.pathreplacing,backgrounds,calc,arrows,arrows.meta,spy,matrix}
\usepgfplotslibrary{patchplots,groupplots}
\usepackage{tikzscale}
\usepackage{balance}

\usepackage{hyperref}
\usepackage[capitalize]{cleveref}
\crefname{section}{Sec.}{Secs.}

\usepackage{subcaption}
\usepackage{graphicx}
\usepackage{caption}
\usetikzlibrary {patterns.meta}
\usetikzlibrary {arrows.meta}

\usepackage{ulem}

\providecolor{added}{rgb}{0,0,1}
\providecolor{deleted}{rgb}{1,0,0}

\makeatletter
\newcommand{\new}[1]{{\textcolor{blue}{#1}}}

\makeatletter
\newcommand\remembertext[2]{
  \immediate\write\@auxout{\unexpanded{\global\long\@namedef{mytext@#1}{#2}}}%
  {\color{blue} #2}%
}
\newcommand\recalltext[1]{%
  \new{\ifcsname mytext@#1\endcsname
    \fontsize{10.5}{12.5}\selectfont\@nameuse{mytext@#1}%
  \else
    ``??''
  \fi
}}
\makeatother

\newcommand{\mpag}[1]{\textcolor{orange}{\textbf{MP: #1}}}

\newacronym{3gpp}{3GPP}{3rd Generation Partnership Project}
\newacronym{adc}{ADC}{Analog to Digital Converter}
\newacronym{5g}{5G}{5th generation}
\newacronym{6g}{6G}{6th generation}
\newacronym{ai}{AI}{Artificial Intelligence}
\newacronym{aimd}{AIMD}{Additive Increase Multiplicative Decrease}
\newacronym{am}{AM}{Acknowledged Mode}
\newacronym{amc}{AMC}{Adaptive Modulation and Coding}
\newacronym{aqm}{AQM}{Active Queue Management}
\newacronym{awgn}{AGWN}{Additive White Gaussian Noise}
\newacronym{balia}{BALIA}{Balanced Link Adaptation}
\newacronym{bdp}{BDP}{Bandwidth-Delay Product}
\newacronym{bf}{BF}{beamforming}
\newacronym{cc}{CC}{Congestion Control}
\newacronym{cdf}{CDF}{Cumulative Distribution Function}
\newacronym{cn}{CN}{Core Network}
\newacronym{cqi}{CQI}{Channel Quality Information}
\newacronym{cp}{CP}{Control Plane}
\newacronym{up}{UP}{User Plane}
\newacronym{upf}{UPF}{User Plane Function}
\newacronym{csirs}{CSI-RS}{Channel State Information - Reference Signal}
\newacronym{dc}{DC}{Dual Connectivity}
\newacronym{rb}{RB}{Resource Block}
\newacronym{dce}{DCE}{Direct Code Execution}
\newacronym{dci}{DCI}{downlink control onformation}
\newacronym{udp}{UDP}{User Datagram Protocol}
\newacronym{dl}{DL}{downlink}
\newacronym{drl}{DRL}{deep reinforcement learning}
\newacronym{fcfs}{FCFS}{first-come-first-served}
\newacronym{dmr}{DMR}{Deadline Miss Ratio}
\newacronym{fspl}{FSPL}{free-space path loss}
\newacronym{dmrs}{DMRS}{DeModulation Reference Signal}
\newacronym{e2e}{E2E}{End-to-End}
\newacronym{ppp}{PPP}{Poission Point Process}
\newacronym{aoi}{AoI}{Area of Interest}
\newacronym{cpu}{CPU}{Central Processing Unit}
\newacronym{lan}{LAN}{Local Area Network}
 \newacronym{gpu}{GPU}{Graphics Processing Unit}
 \newacronym{tpu}{TPU}{Tensor Processing Unit}
\newacronym{si}{SI}{Study Item}
\newacronym{ecn}{ECN}{Explicit Congestion Notification}
\newacronym{edf}{EDF}{Earliest Deadline First}
\newacronym{enb}{eNB}{eNodeB}
\newacronym{epc}{EPC}{Evolved Packet Core}
\newacronym{es}{ES}{Edge Server}
\newacronym{cav}{CAV}{Connected and Autonomous Vehicle}
\newacronym{fdma}{FDMA}{Frequency Division Multiple Access}
\newacronym{fdd}{FDD}{Frequency Division Duplexing}
\newacronym{fr2}{FR2}{Frequency Range 2}
\newacronym{fr1}{FR1}{Frequency Range 1}
\newacronym{tdm}{TDM}{Time Division Multiplexing}
\newacronym{upa}{UPA}{Uniform Planar Array}
\newacronym{car}{CAR}{Circular Aperture Reflector }
\newacronym[firstplural=Radio Access Technologies (RATs)]{rat}{RAT}{Radio Access Technology}
\newacronym[firstplural=Radio Access Technology (RTs)]{rt}{RT}{Radio Technology}
\newacronym{fs}{FS}{Fast Switching}
\newacronym{isd}{ISD}{inter-site distance}
\newacronym{ftp}{FTP}{File Transfer Protocol}
\newacronym{gnb}{gNB}{Next Generation Node Base}
\newacronym{harq}{HARQ}{Hybrid Automatic Repeat reQuest}
\newacronym{hetnet}{HetNet}{Heterogeneous Network}
\newacronym{hh}{HH}{Hard Handover}
\newacronym{hol}{HOL}{Head-of-Line}
\newacronym{ia}{IA}{Initial Access}
\newacronym{imt}{IMT}{International Mobile Telecommunication}
\newacronym{iot}{IoT}{Internet of Things}
\newacronym{los}{LOS}{Line of Sight}
\newacronym{lte}{LTE}{Long Term Evolution}
\newacronym{m2m}{M2M}{Machine to Machine}
\newacronym{mac}{MAC}{Medium Access Control}
\newacronym{mc}{MC}{Multi-Connectivity}
\newacronym{mcs}{MCS}{Modulation and Coding Scheme}
\newacronym{mec}{MEC}{Mobile Edge Cloud}
\newacronym{mi}{MI}{Mutual Information}
\newacronym{mimo}{MIMO}{Multiple Input Multiple Output}
\newacronym{milp}{MILP}{Mixed-Integer Linear Programming}
\newacronym{mmwave}{mmWave}{millimeter wave}
\newacronym{mptcp}{MPTCP}{Multipath TCP}
\newacronym{mr}{MR}{Maximum Rate}
\newacronym{mss}{MSS}{Maximum Segment Size}
\newacronym{mtd}{MTD}{Machine-Type Device}
\newacronym{mtu}{MTU}{Maximum Transmission Unit}
\newacronym{nfv}{NFV}{Network Function Virtualization}
\newacronym{vnf}{VNF}{Virtualization Network Function}
\newacronym{gv}{GV}{ground vehicle}
\newacronym{vec}{VEC}{Vehicular Edge Computing}
\newacronym{dn}{DN}{Data Network}
\newacronym{sdn}{SDN}{Software Defined Networking}
\newacronym{nlos}{NLOS}{Non Line of Sight}
\newacronym{nlosb}{NLOSb}{Building Non Line of Sight}
\newacronym{nlosv}{NLOSv}{Vehicle Non Line of Sight}
\newacronym{nr}{NR}{New Radio}
\newacronym{ofdm}{OFDM}{Orthogonal Frequency Division Multiplexing}
\newacronym{pdcch}{PDCCH}{Physical Downlonk Control Channel}
\newacronym{sctp}{SCTP}{Stream Control Transport Protocol}
\newacronym{sdap}{SDAP}{Service Data Adaptation Protocol}
\newacronym{pdcp}{PDCP}{Packet Data Convergence Protocol}
\newacronym{pdsch}{PDSCH}{Physical Downlink Shared Channel}
\newacronym{pdu}{PDU}{Packet Data Unit}
\newacronym{sdu}{SDU}{Service Data Unit}
\newacronym{pf}{PF}{Proportional Fair}
\newacronym{pgw}{PGW}{Packet Gateway}
\newacronym{sgw}{SGW}{Serving Gateway}
\newacronym{phy}{PHY}{Physical}
\newacronym{pbch}{PBCH}{Physical Broadcast Channel}
\newacronym[plural=\gls{mme}s,firstplural=Mobility Management Entities (MMEs)]{mme}{MME}{Mobility Management Entity}
\newacronym{prb}{PRB}{Physical Resource Block}
\newacronym{pss}{PSS}{Primary Synchronization Signal}
\newacronym{pucch}{PUCCH}{Physical Uplink Control Channel}
\newacronym{pusch}{PUSCH}{Physical Uplink Shared Channel}
\newacronym{rach}{RACH}{Random Access Channel}
\newacronym{ran}{RAN}{Radio Access Network}
\newacronym{red}{RED}{Random Early Detection}
\newacronym{rf}{RF}{Radio Frequency}
\newacronym{rlc}{RLC}{Radio Link Control}
\newacronym{rlf}{RLF}{Radio Link Failure}
\newacronym{rrc}{RRC}{Radio Resource Control}
\newacronym{rrm}{RRM}{Radio Resource Management}
\newacronym{rr}{RR}{Round Robin}
\newacronym{rs}{RS}{Remote Server}
\newacronym{rsrp}{RSRP}{Reference Signal Received Power}
\newacronym{rss}{RSS}{Received Signal Strength}
\newacronym{rtt}{RTT}{Round Trip Time}
\newacronym{rw}{RW}{Receive Window}
\newacronym{rx}{RX}{Receiver}
\newacronym{sa}{SA}{standalone}
\newacronym{sack}{SACK}{Selective Acknowledgment}
\newacronym{sap}{SAP}{Service Access Point}
\newacronym{sch}{SCH}{Secondary Cell Handover}
\newacronym{scoot}{SCOOT}{Split Cycle Offset Optimization Technique}
\newacronym{sdma}{SDMA}{Spatial Division Multiple Access}
\newacronym{sinr}{SINR}{Signal to Interference plus Noise Ratio}
\newacronym{sm}{SM}{Saturation Mode}
\newacronym{snr}{SNR}{Signal-to-Noise Ratio}
\newacronym{son}{SON}{Self-Organizing Network}
\newacronym{ss}{SS}{Synchronization Signal}
\newacronym{srs}{SRS}{Sounding Reference Signal}
\newacronym{sss}{SSS}{Secondary Synchronization Signal}
\newacronym{tb}{TB}{Transport Block}
\newacronym{tcp}{TCP}{Transmission Control Protocol}
\newacronym{tdd}{TDD}{Time Division Duplexing}
\newacronym{tdma}{TDMA}{Time Division Multiple Access}
\newacronym{tfl}{TfL}{Transport for London}
\newacronym{tm}{TM}{Transparent Mode}
\newacronym{prr}{PRR}{Packet Reception Ratio}
\newacronym{trp}{TRP}{Transmitter Receiver Pair}
\newacronym{tti}{TTI}{Transmission Time Interval}
\newacronym{ttt}{TTT}{Time-to-Trigger}
\newacronym{tx}{TX}{Transmitter}
\newacronym{ue}{UE}{User Equipment}
\newacronym{ul}{UL}{uplink}
\newacronym{uml}{UML}{Unified Modeling Language}
\newacronym{um}{UM}{Unacknowledged Mode}
\newacronym{utc}{UTC}{Urban Traffic Control}
\newacronym{vm}{VM}{Virtual Machine}
\newacronym{rsrq}{RSRQ}{Reference Signal Received Quality}
\newacronym{rssi}{RSSI}{Received Signal Strength Indicator}
\newacronym{crs}{CRS}{Cell Reference Signal}
\newacronym{v2v}{V2V}{Vehicle-to-Vehicle}
\newacronym{v2i}{V2I}{Vehicle-to-Infrastructure}
\newacronym{v2n}{V2N}{Vehicle-to-Network}
\newacronym{v2x}{V2X}{Vehicle-to-Everything}
\newacronym{vn}{VN}{Vehicular Node}
\newacronym{dsrc}{DSRC}{Dedicated Short Range Communication}
\newacronym{ci}{CI}{context information}
\newacronym{voi}{VoI}{value of information}
\newacronym{gps}{GPS}{Global Positioning System}
\newacronym{qos}{QoS}{Quality of Service}
\newacronym{qoe}{QoE}{Quality of Experience}
\newacronym{ml}{ML}{Machine Learning}
\newacronym{ahp}{AHP}{Analytic Hierarchy Process}
\newacronym{lidar}{LIDAR}{Light Detection and Ranging}
\newacronym{sumo}{SUMO}{Simulation of Urban MObility}
\newacronym{wave}{WAVE}{Wireless Access in Vehicular Environment}
\newacronym{c-its}{C-ITS}{Connected Intelligent Transportation System}
\newacronym{dash}{DASH}{Dynamic Adaptive Streaming over HTTP}
\newacronym{http}{HTTP}{HyperText Transfer Protocol}
\newacronym{nt}{NT}{Non-Terrestrial}
\newacronym{ntc}{NTC}{non-terrestrial communication}
\newacronym{ntn}{NTN}{Non-Terrestrial Network}
\newacronym{haps}{HAPS}{High Altitude Platform Station}
\newacronym{hap}{HAP}{High Altitude Platform}
\newacronym{leo}{LEO}{Low Earth Orbit}
\newacronym{meo}{MEO}{Medium Earth Orbit}
\newacronym{geo}{GEO}{Geostationary Earth Orbit}
\newacronym{uav}{UAV}{unmanned aerial vehicle}
\newacronym{nsat}{nSAT}{Nanosatellite}
\newacronym{ehf}{EHF}{extremely high-frequency}
\newacronym{ioe}{IoE}{Internet of Everyone}
\newacronym{gan}{GaN}{Gallium Nitride}
\newacronym{af}{AF}{amplify-and-forward}
\newacronym{csi}{CSI}{channel state information}
\newacronym{ecdf}{ECDF}{empirical cumulative distribution function}
\newacronym{f}{F}{flexible}
\newacronym{fpga}{FPGA}{field programmable gate array}
\newacronym{fov}{FoV}{field-of-view}
\newacronym{km}{KM}{K-means}
\newacronym{kmed}{KMed}{K-medoids}
\newacronym{iab}{IAB}{Integrated Access and Backhaul}
\newacronym{bap}{BAP}{backhaul adaptation protocol}
\newacronym{irs}{IRS}{intelligent reflecting surface}
\newacronym{lsfc}{LSFC}{large-scale fading coefficient}
\newacronym{noma}{NOMA}{non-orthogonal multiple access}
\newacronym{fdm}{FDM}{Frequency Division Multiplexing}
\newacronym{sdm}{SDM}{space-division multiplexing}
\newacronym{ofdma}{OFDMA}{orthogonal frequency-division multiple access}
\newacronym{oma}{OMA}{orthogonal multiple access}
\newacronym{plos}{pLoS}{probabilistic \ac{los}}
\newacronym{rsma}{RSMA}{rate-splitting multiple access}
\newacronym{scm}{SCM}{spatial channel model}
\newacronym{siso}{SISO}{single input single output}
\newacronym{svd}{SVD}{singular value decomposition}
\newacronym{thz}{THz}{Terahertz}
\newacronym{ula}{ULA}{uniform linear array}
\newacronym{uma}{UMa}{urban macro-cell}
\newacronym{umi}{UMi}{urban micro-cell}
\newacronym{mt}{MT}{mobile terminal}
\newacronym{cu}{CU}{centralized unit}
\newacronym{du}{DU}{distributed unit}
\newacronym{dag}{DAG}{directed acyclic graph}
\newacronym{st}{ST}{spanning tree}
\newacronym{rma}{RMa}{rural macrocell}
\newacronym{inf}{InF}{indoor factory}
\newacronym{ngc}{NGC}{Next Generation Core}
\newacronym{gtp}{GTP}{GPRS Tunnelling Protocol}
\newacronym{tft}{TFT}{Traffic Flow Template}
\newacronym{teid}{TEID}{Tunnel Endpoint Identifier}
\newacronym{tnl}{TNL}{Transport Network Layer}
\newacronym{amf}{AMF}{Access and Mobility Management Function}
\newacronym{pdr}{PDR}{Packet Delivery Ratio}
\newacronym{embb}{eMBB}{enhanced Mobile Broadband}
\newacronym{5gc}{5GC}{5G Core}
\newacronym{oran}{O-RAN}{Open Radio Access Network}
\newacronym{dscp}{DSCP}{Differentiated Services Code Point}
\newacronym{m-iab}{mIAB}{mobile Integrated Access and Backhaul}
\newacronym{its}{ITS}{Intelligent Transportation Systems}

\title{End-to-End Simulation of 5G NR \\ Integrated Access and Backhaul Networks \\ for Remote Maritime Connectivity}

\author{Alessandro Traspadini, Matteo Pagin, Rapha\"{e}l Ihamouine, Rupert Lucas, \\ Andrew Noren, Michele Zorzi, \textit{Fellow, IEEE}, Marco Giordani, \textit{Senior Member, IEEE} 
\vspace{-2em}
\thanks{
A. Traspadini, M. Zorzi, and M. Giordani are with the Department of Information Engineering, University of Padova. Padova, Italy. (E-mail: \{alessandro.traspadini, marco.giordani, michele.zorzi\}@dei.unipd.it).\\
M. Pagin was with the Department of Information Engineering, University of Padova. Padova, Italy. He is now with Keysight Denmark.\\
R. Ihamouine, R. Lucas, and A. Noren are with Viasat Inc, UK. (Email: \{Raphael.Ihamouine, Rupert.Lucas, Andrew.Noren\}@viasat.com).\\
This work was partially supported by the European Union under the Italian National Recovery and Resilience Plan (NRRP) Mission 4, Component 2, Investment 1.3, CUP C93C22005250001, partnership on “Telecommunications of the Future” (PE00000001 - program “RESTART”).}}

\begin{document}
\bstctlcite{IEEEexample:BSTcontrol}
\maketitle

\begin{abstract}
\Gls{mmwave} \gls{5g} networks offer high data rates but face coverage challenges due to severe path loss and blockage.
These problems motivate the use of \gls{iab} as a flexible wireless backhaul solution that extends connectivity to cell boundaries and unfibered areas, including maritime environments.
This paper overviews the latest 3GPP specifications for IAB networks in Releases 16 through 18. Then, it presents an ns-3 module for IAB, featuring a complete end-to-end protocol stack, including the \gls{bap} layer, flexible slot and control configurations, and multiplexing schemes based on both time and frequency division.
We test the IAB module via extensive system-level simulations in a custom maritime scenario 
where vessels, equipped with \gls{iab}-nodes, can simultaneously act as access points and relays, forming dynamic multi-hop networks that maintain connectivity via wireless backhaul to shore-based stations.
We evaluate different topologies and channel conditions, providing insights into the design and deployment of \gls{mmwave} \gls{iab} networks in offshore environments.
\end{abstract}


\begin{tikzpicture}[remember picture,overlay]
\node[anchor=north,yshift=-10pt] at (current page.north) {\parbox{\dimexpr\textwidth-\fboxsep-\fboxrule\relax}{
\centering\footnotesize This paper has been accepted for publication at IEEE Transactions on Communications, 2026.

Please cite it as: 
A. Traspadini, M. Pagin, R. Ihamouine, R. Lucas, A. Noren, M. Zorzi, and M. Giordani, 
"End-to-End Simulation of 5G NR Integrated Access and Backhaul Networks for Remote Maritime Connectivity," 
IEEE Transactions on Communications, to appear, 2026.}};
\end{tikzpicture}

\glsresetall

\section{Introduction}
\label{sec:introduction}

Current \gls{5g} \gls{nr} commercial networks are designed to operate in both \gls{fr1} below 6~GHz and \gls{fr2} in the \gls{mmwave} spectrum~\cite{Tafintsev2024}.
The \gls{mmwave} bands, in particular, enable gigabit-level data rates for \gls{embb} services, but face significant propagation challenges due to severe path loss, blockage, and diffuse scattering. To ensure ubiquitous and continuous coverage, networks should be deployed as dense small cells, to reduce inter-site distance and establish stronger access channels~\cite{lopez2015towards}.
This approach, however, involves high capital and operational expenditures (capex and opex) for network operators, primarily due to the high costs and logistical complexity associated with the installation of fiber-based backhaul.

To address these challenges, the \gls{3gpp} standardized \gls{iab} in Release 16~\cite{3gpp_38_874}, as a cost-efficient solution to support dense \gls{5g} deployments without requiring fiber connectivity at every site.
In IAB, only a fraction of \glspl{gnb}, called IAB-donors, are directly connected to the \gls{5gc} via traditional fiber links. The others, called IAB-nodes, simultaneously operate as access points for \glspl{ue} and as wireless relays that forward the backhaul traffic toward the IAB-donor, possibly via multiple hops and at mmWave frequencies to maximize capacity~\cite{polese2020integrated, Zhang2021IABSurvey}.
IAB provides a flexible and easy-to-deploy solution to improve coverage and accessibility, making it an attractive option for mmWave \gls{5g} networks~\cite{Sadovaya2022, Topcu2025}. 
In fact, several operators have already shown interest in deploying \gls{iab}, with current estimates suggesting that approximately 10–20\% of \gls{5g} sites could adopt this technology~\cite{Tafintsev2024}.


\gls{iab} is particularly beneficial in those scenarios where \gls{los} propagation is blocked, such as in urban environments, or where fiber backhaul is impractical, such as in rural and remote areas~\cite{Sadovaya2022, Topcu2025}, as well as for \gls{its}, including railway and vehicular networks~\cite{Monteiro22Paving, Monteiro22Tdd,Choi2021,Li2022}.
Beyond terrestrial applications, IAB can also play a crucial role to extend coverage in remote maritime coastal areas and seaports. In this context, \gls{iab}-nodes can be installed on vessels to form a dynamic multi-hop access and backhaul network interconnecting ships and shore stations with no or limited fiber capabilities~\cite{Horsmanheimo2024,Capela2021}.
When sidelink communication is enabled, IAB can also support direct ship-to-ship connectivity for exchanging critical information, such as navigation data and safety alerts, even beyond cellular coverage.
Recent works have highlighted the role of sidelink IAB to enable autonomous maritime operations, e.g., to coordinate maneuvers among fleets of ships with low latency~\cite{Rafiq2025}.

To properly design, optimize, and dimension IAB networks, accurate and realistic end-to-end evaluation is required.
In this context, ns-3 enables full-stack simulation of complex network scenarios, and therefore represents a valid research tool in this domain.
In 2018, we released an open-source \gls{5g} \gls{iab} simulation module for ns-3~\cite{polese2018end}.
However, that module was based on early 3GPP specifications and assumptions. Since then, the standardization of IAB has progressed significantly, and several key features have been introduced from Release 16.
New features include: (i) a new \gls{bap} layer, located above the \gls{rlc} layer, to support routing and forwarding of the packets across the IAB topology, i.e., from the
IAB-donor to the access IAB-node; (ii) the support for \gls{dc}, which permits an IAB-node
 to concurrently connect to multiple IAB-donors; (iii) new multiplexing and scheduling mechanisms; (iv) improved topology
adaptation, interference mitigation, and mobility management
schemes to support \gls{m-iab} nodes and UEs in dynamic environments.
These innovations necessitate substantial updates of the existing ns-3 module to reflect current IAB research~trends.

Based on the above introduction, in this paper we provide the following contributions:
\begin{itemize}
\item We present a comprehensive overview of the main \gls{3gpp} \gls{5g} \gls{nr} \gls{iab} specifications from Releases 16 to 18. 

    \item We design and implement an open-source system-level simulation framework for \gls{5g} \gls{nr} \gls{iab} networks based on ns-3, called \texttt{ns3-mmwave-iab}, which incorporates realistic physical layer modeling.
    Compared to the earlier implementation of this module presented in~\cite{polese2018end}, the proposed framework is now aligned with the latest \gls{3gpp} IAB specifications.
    The framework captures realistic slot structures, multiplexing schemes (supporting both \gls{tdm} and \gls{fdm}), configurable control overhead, and multi-hop backhaul constraints.
    
    \item Most of the literature on \gls{iab} is based on terrestrial networks, while the maritime scenario remains largely unexplored. 
    However, emerging trends in commercial, military, security, safety-critical, navigation, and offshore monitoring operations rely on continuous and high-capacity connectivity, and highlight the strategic role of \gls{iab} in maritime scenarios.
    To address this gap, our \texttt{ns3-mmwave-iab} framework also integrates a maritime-specific \gls{mmwave} propagation and channel model, to evaluate the effects of sea-surface reflections and offshore propagation characteristics.
    
    \item Leveraging this framework, we conduct an extensive simulation campaign to evaluate the performance of maritime \gls{iab} networks under different topologies, weather, and resource allocation configurations.
    We demonstrate that single-hop deployments offer low latency in good propagation conditions, but suffer from severe rain attenuation. In turn, multi-hop networks suffer more interference, though this impact can be mitigated by heavy rain.
    Resource allocation is also crucial: in a multi-hop \gls{iab} network, we discuss how to distribute \gls{ofdm} symbols to avoid bottleneck effects.
\end{itemize}


The remainder of this paper is organized as follows.
\Cref{sec:sota} reviews the related work on \gls{iab}. 
\Cref{sec:iab-standard} outlines the 3GPP \gls{iab} standard from Release~16 to Release~18.
\Cref{sec:simulation-module} describes the proposed \gls{iab} ns-3 simulator, focusing on the implemented protocol stack, maritime channel model, slot formats, control symbols, and multiplexing.
\Cref{sec:perf-eval} presents our main simulation results.
Finally, \cref{sec:conclusions} concludes the paper with directions for future research.

\section{Related Works}
\label{sec:sota}
The existing literature on \gls{iab} has investigated key research aspects such as resource allocation, topology optimization, and mobility support, under both static and dynamic conditions, which are reviewed in the following paragraphs.

\paragraph{Resource allocation}
\gls{iab}-nodes can operate in both in-band and out-of-band scenarios.
The out-of-band mode involves using different bands for access and backhaul, while in the in-band mode the backhaul and access links share a common band, which provides more flexibility at the cost of increased complexity~\cite{Monteiro22Paving}.
Resource allocation has primarily focused on in-band backhauling networks~\cite{Tafintsev2024, Zhang20Joint, Lai20Resource, Pagin22Resource, Liu20Joint, Saha19Millimeter, Lei20Deep,Tian19Field}, while fewer works, e.g.,~\cite{Topcu2025,Kwon22OutBand}, have addressed out-of-band backhauling due to the higher complexity.

In~\cite{Tafintsev2024}, the authors proposed a joint resource allocation and link scheduling framework for multi-hop \gls{mmwave} \gls{iab} systems, considering half-duplex constraints and flexible \gls{tti} configurations, showing that multiple beams at the \gls{iab}-donor can significantly improve the throughput.
Similarly, \cite{Zhang20Joint} investigated capacity maximization in a multi-hop in-band \gls{iab} network using approximation techniques to solve the underlying optimization problem.

Centralized and semi-centralized resource allocation strategies were explored in~\cite{Lai20Resource,Pagin22Resource}, showing that centralized decisions can improve the performance but are sensitive to outdated channel quality information, e.g., \gls{csi}, obtained from child nodes.
To address this limitation, semi-centralized resource allocation allows local schedulers to refine centralized decisions based on local information, provided that timely information exchange is feasible, thereby improving the network performance.

Similarly, in~\cite{Saha19Millimeter} the authors proposed an integrated resource allocation scheme, where the bandwidth assigned by a donor \gls{iab} to the backhaul for a given \gls{iab}-node depends on the access load of the node itself. 
This scheme can be interpreted as a centralized allocation framework with some adaptation to local conditions, and provides superior coverage probability compared to static resource allocation.
Alternative approaches include game-theoretic formulations~\cite{Liu20Joint}, which optimize resource allocation via centralized and distributed algorithms converging to the Nash bargaining solution, and methods based on \gls{drl}~\cite{Lei20Deep}, which adaptively allocate spectrum in dynamic \gls{iab} networks while handling time-varying conditions and heterogeneous setups.
An experimental approach was presented in~\cite{Tian19Field}, where the authors implemented a testbed consisting of an \gls{iab}-donor, an \gls{iab}-node, and a user operating in the \gls{mmwave} band, and evaluated dynamic resource allocation between access and backhaul links in terms of coverage and throughput.

In the out-of-band context, the authors in~\cite{Topcu2025} proposed an \gls{oran} optimization framework for out-of-band \gls{iab}, formulating a \gls{milp} to jointly optimize user-level radio resources, bandwidth partitioning, and \gls{mmwave} backhaul routing.
Similarly, out-of-band \gls{mmwave} \gls{iab} networks were studied in~\cite{Kwon22OutBand} with both centralized and distributed resource allocation algorithms, complemented by a geometric analysis to approximate co-channel interference and provide system design insights.


\begin{figure*}[t!]
    \centering
\includegraphics[width=.9\textwidth]{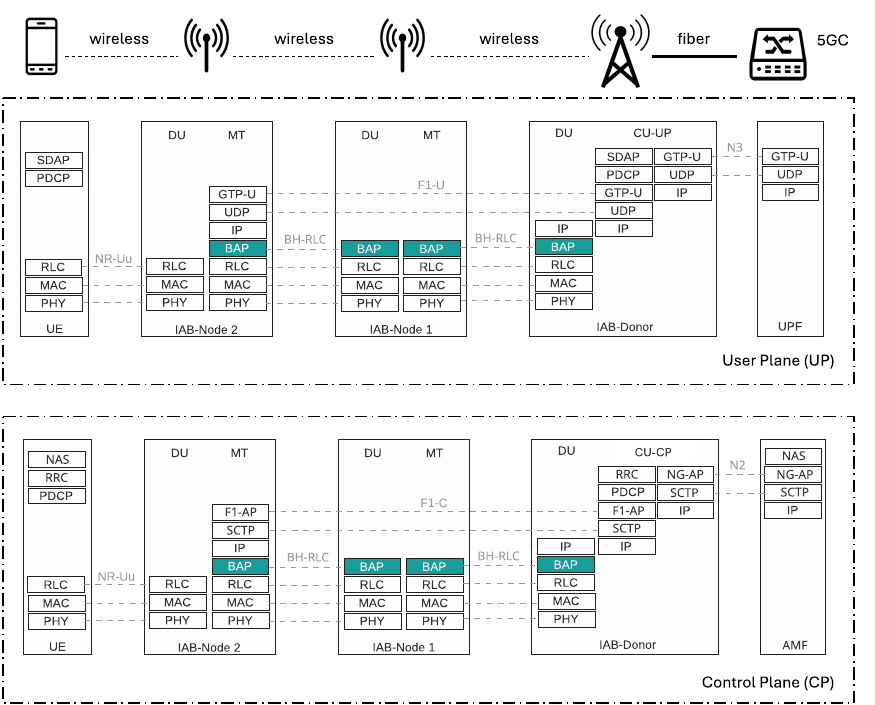}
   \caption{3GPP IAB protocol stacks for UP and CP. 
    In both cases, packets are routed from the IAB-Donor to IAB-Node 2 through the wireless backhaul of IAB-Node 1.}
    \label{fig:iab-stack-3gpp}
    \vskip -0.1em
\end{figure*}

\paragraph{Topology optimization}

Topology design and routing strategies are fundamental to achieve efficient multi-hop communication.
Several works have addressed topology optimization using centralized and distributed approaches~\cite{Alghafari22Decentralized,Fang21Joint,Polese18Distributed,Islam18Investigation}.

In~\cite{Alghafari22Decentralized}, the authors proposed a distributed stochastic optimization framework for joint resource allocation and path selection.
Similarly, \cite{Fang21Joint} investigated joint routing and resource allocation to maximize the minimum node throughput under time and resource constraints in a \gls{tdma} multi-hop \gls{iab} network.
Distributed routing strategies (in which each \gls{iab}-node makes the next-hop decision) were studied in~\cite{Polese18Distributed}, where different path selection strategies were compared, including those based on the highest \gls{sinr}, the shortest path, and the load conditions on the routes. 
The authors showed that the shortest path strategy, which always selects an \gls{iab}-donor as the next hop, if reachable, is ineffective in sparsely deployed networks and for low \gls{sinr} regimes.

Mesh and tree-based IAB topologies were compared in~\cite{Islam18Investigation}, showing that mesh deployments can improve throughput and latency due to increased path diversity.
Nevertheless, these studies focused primarily on algorithmic optimization, and do not capture the full protocol stack or realistic physical layer~constraints.

\paragraph{Mobility}

Mobility introduces additional challenges due to dynamic topology changes and varying channel conditions. To mitigate interference in \gls{m-iab} networks, silent slots in the \gls{tdd} frame structure were proposed in~\cite{Monteiro22Tdd}.  
The integration of \glspl{uav} as \gls{m-iab}-nodes has also been widely studied~\cite{Yaliniz195G, Zhang19Framework, Gapeyenko18Flexible}, showing that aerial relays can enhance coverage and enable flexible network deployment.
For instance, \cite{Zhang19Framework} proposed joint optimization of power allocation, resource allocation, and \gls{uav} positioning to improve throughput, while \cite{Gapeyenko18Flexible} explored dynamic rerouting strategies using mathematical models to maintain connectivity in the presence of link~blockages.

\section{IAB in the 3GPP}
\label{sec:iab-standard}

In this section we review the 3GPP \gls{iab} specifications from Release 16.
\cref{sec:3gpp_arch} introduces the \gls{iab} architecture, \cref{sec:3gpp_bap} describes the \gls{bap}, \cref{sec:3gpp_top} presents the supported \gls{iab} topologies, \cref{sec:3gpp_mul} discusses multiplexing and scheduling mechanisms, and \cref{sec:3gpp_mob} focuses on mobility management enhancements introduced in Release 18.

\subsection{IAB Architecture}
\label{sec:3gpp_arch}
IAB was standardized in the \gls{3gpp} 5G NR specifications in Release 16~\cite{3gpp_38_874}, and is based on the functional split paradigm introduced in Release 15.
In this architecture, illustrated in~\cref{fig:iab-stack-3gpp}, we distinguish three types of nodes:
\begin{itemize}
    \item \glspl{ue}, acting as end users. The Uu interface connects the UE and its parent IAB-node.
    \item IAB-nodes, i.e., \glspl{gnb} with wireless backhaul.
    Each IAB-node is divided into two logical units: a \gls{mt}, for the
wireless backhaul connection toward an upstream IAB-node or IAB-donor, and a \gls{du}, for the access connection to the UEs or the downstream MTs of other IAB-nodes. 
    \item IAB-donors, i.e., full-blown \glspl{gnb} with a classical wired (fiber) connection toward the core network. Each IAB-donor is also divided into two logical units: a \gls{cu} and a \gls{du}. Specifically, the \gls{du} is closer to the radio and antenna elements, and implements a limited subset of time-critical radio functions such as scheduling, segmentation, and packet retransmission~\cite{Madapatha20Iab}.
\end{itemize}

According to ``Option 2'' in TR 38.801~\cite{3gpp.38.801}, the \gls{du} only terminates the \gls{rlc}, \gls{mac}, and \gls{phy} layers.
For the \gls{up}, the \gls{cu}-\gls{up} of the \gls{iab}-donor and the UE terminate the \gls{sdap} and \gls{pdcp} layers. Data packets are exchanged via the F1-U interface, which consists of \gls{gtp}-U tunnels from the \gls{cu}-UP and the serving \gls{iab}-node \gls{du}.
For the \gls{cp}, the \gls{cu}-\gls{cp} of the \gls{iab}-donor and the UE terminate the \gls{rrc} and \gls{pdcp} layers. Control packets are exchanged via the F1-C interface, which uses the \gls{sctp} between the CU-CP and the serving \gls{iab}-node \gls{du}.

\subsection{Backhaul Adaptation Protocol (BAP)}
\label{sec:3gpp_bap}

The \gls{bap} is an additional layer located above the \gls{rlc}, as depicted in~\cref{fig:iab-stack-3gpp}, which was introduced by the \gls{3gpp} to route the packets across the IAB topology, i.e., 
from the \gls{iab}-donor to the access \gls{iab}-node~\cite{38340}.
Each \gls{iab}-node is assigned a unique \gls{bap} address by the \gls{iab}-donor. For \gls{dl} packets, the \gls{bap} layer of the \gls{iab}-donor adds a \gls{bap} header that includes the destination \gls{bap} address of the access \gls{iab}-node and the path ID, to discriminate among different routes to the destination.
Similarly, for \gls{ul} packets, the access \gls{iab}-node adds a \gls{bap} header containing the destination \gls{bap} address of the \gls{iab}-donor and the path ID.
Each \gls{iab}-node is configured with routing tables for \gls{ul} and \gls{dl}, indicating the child node (in the case of \gls{dl}) or parent node (in the case of \gls{ul}) to which the packet should be forwarded. 
Upon receiving a packet, the \gls{bap} layer checks the destination address in the \gls{bap} header.
If it matches its own address, the packet is forwarded to the higher layers of the \gls{du}. Otherwise, the \gls{iab}-node delivers the packet to its \gls{du} and forwards it to the next node based on the routing table.

The \gls{bap} layer is also responsible for mapping ingress and egress backhaul \gls{rlc} channels to ensure that packets meet the desired \gls{qos} requirements.
For bearers with strict \gls{qos} demands, a $1:1$ mapping can be used, to dedicate an individual backhaul \gls{rlc} channel to each hop.
Otherwise, a $1:N$ mapping is employed to multiplex packets from $N$ bearers over a single backhaul \gls{rlc} channel~\cite{Madapatha20Iab}.

\subsection{IAB Topologies}
\label{sec:3gpp_top}
\gls{iab} deployments exhibit a \textit{possibly multi-hop} topology, where a well-defined parent-child relationship is present. Parent nodes can be represented by either an \gls{iab}-donor or an \gls{iab}-node; child nodes by either \glspl{ue} or downstream \gls{iab}-nodes.
In 3GPP 5G NR Release 16, the standard supported two types of \gls{iab} topologies~\cite{3gpp_38_874}: \gls{st} and \gls{dag}. 
In the former, \gls{iab}-nodes are connected to a single parent, 
while in the latter multiple child-to-parent connections can be established.
Clearly, an \gls{st} topology is less complex, but at the same time comes with resiliency constraints and performance degradation.
For instance, in an \gls{st}, backhaul 
 \glspl{rlf} are likely to result in service interruptions for the end users, due to the lack of alternative backhaul routes.
Instead, the \gls{dag} topology provides backhaul route redundancy, which can be used both to increase service availability and for load balancing.
Notably, the \gls{3gpp} does not set an upper limit on the depth of either topology. Therefore, \gls{iab} protocols must provide support for an arbitrary number of backhaul hops~\cite{3gpp_38_874}.

Since Release 17, the 3GPP also supports an IAB-node \gls{mt} to concurrently connect to two IAB-donors via NR \gls{dc}~\cite[Sec. 11.3]{3gpp.21.917}. 
This approach effectively interconnects two IAB-topologies, with the dual-connected IAB-node taking the role of the boundary node. Accordingly, it is in charge of possibly overwriting the BAP header routing information whenever packets are routed across the two~topologies.

\subsection{Multiplexing and Scheduling}
\label{sec:3gpp_mul}
In an IAB network, the backhaul link between a parent node
DU and the child node \gls{mt} mimics the link between a fiber-equipped gNB and a typical 5G NR UE.
Accordingly, there exist three possible types of time-domain resources just like for 5G NR \glspl{ue}, i.e., \gls{dl}, \gls{ul}, or \gls{f}~\cite[Sec. 11.1]{3gpp.38.213}.
However, in practice, the actual transmission direction (UL or DL) depends on the current availability of resources, and the IAB configuration, i.e., half- or full-duplex.
Specifically, in case of half-duplex (when ``IAB Simultaneous Operation'' is disabled~\cite{3gpp.38.174}), DUs/MTs cannot concurrently transmit and receive data.
To reflect this constraint, time resources can also be marked as~\cite{3gpp.R1.1913600}:
\begin{itemize}
    \item {Hard.} 
    The resource is always available to the DU, regardless of the MT configuration. Accordingly, other half-duplex MTs will mark these resources as ``Not~Available''.
    \item {Soft.} 
    The resource is available to the DU if and only if it does not collide with other MT transmissions/receptions. When ``{IAB Simultaneous Operation}'' is enabled, this is equivalent to a ``{Hard}'' resource. 
    The actual availability of ``{Soft}'' resources can be inferred either implicitly or explicitly (in the latter case, based on the current resource allocations of the MTs).
        \item {Not Available.} 
    The resource is unavailable due to half-duplexing constraints.
\end{itemize}

\begin{figure*}[ht!]
    \centering
    \begin{subfigure}[b]{0.8\textwidth}
        \centering
        \includegraphics[width=\textwidth]{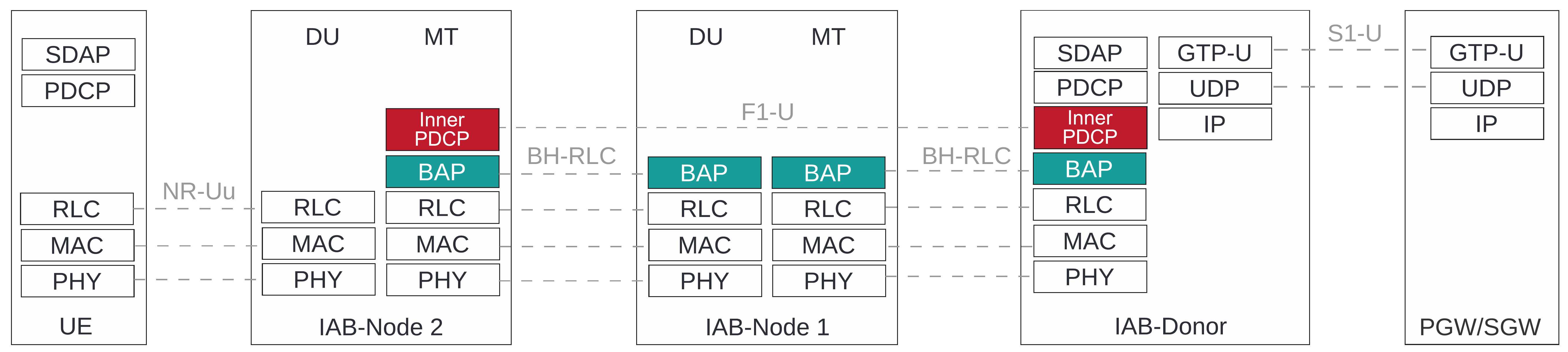}
        \caption{PDU session traffic (UE-anchored).}
        \vspace{0.4cm}
        \label{fig:iab-stack-sim-pdu}
    \end{subfigure}
    \begin{subfigure}[b]{0.7\textwidth}
        \centering
        \includegraphics[width=\textwidth]{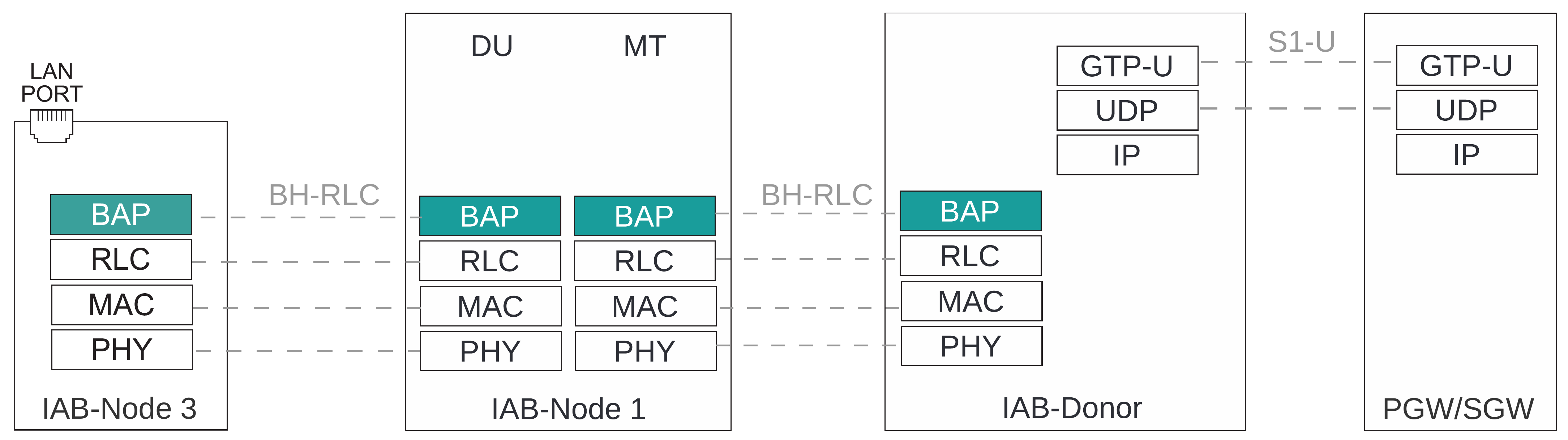}
        \caption{Non-PDU traffic (LAN-anchored).}
        \vspace{0.4cm}
        \label{fig:iab-stack-sim-lan}
    \end{subfigure}
    \caption{User-plane protocol stacks implemented in our IAB simulator.
    In \cref{fig:iab-stack-sim-pdu}, PDU session traffic is anchored at the IAB-donor CU-UP, and uses the ``inner PDCP'' entity to establish internal GTP-U tunnels over the IAB backhaul.
    In \cref{fig:iab-stack-sim-lan}, non-PDU traffic bypasses the SDAP, PDCP, and ``inner PDCP'' layers, and is forwarded across the IAB backhaul via BAP-only routing.}
    \label{fig:iab-stack-sim}
    \vskip -0.1em
\end{figure*}


Since Release 17, the 3GPP also supports \gls{fdm} of MTs and DUs by marking resources as ``{Hard},'' ``{Soft},'' and ``{Not Available}'' on a per-RB basis. This approach effectively enables concurrent transmission/reception by the \gls{mt} and \gls{du} interfaces, utilizing frequency division and spatial multiplexing techniques~\cite[Sec. 11.3]{3gpp.21.917}. Moreover, Release 17 further promotes \gls{sdm} in IAB networks by introducing signaling between neighboring IAB-nodes regarding restricted and/or preferred beams, power adjustments, and transmission/reception timing alignments~\cite[Sec. 11.3]{3gpp.21.917}.


\subsection{Mobility Management}
\label{sec:3gpp_mob}

3GPP Release 18 introduces innovations for topology adaptation, interference mitigation, and mobility management to support \gls{iab}-nodes and \glspl{ue} in dynamic environments~\cite{lin22advanced}.

Focusing on the latter aspect, a key improvement is the management of \gls{m-iab}-nodes, such as those deployed in moving vehicles like buses or trains~\cite{wen24shaping}.
These nodes can provide \gls{5g} connectivity, but their mobility introduces challenges during \gls{iab}-donor switching, potentially causing disruptions for the \glspl{ue}.
To address this issue, Release 18 introduces a dedicated mobile \gls{cu} (m-CU) with an Xn interface that connects to multiple \gls{iab}-donors~\cite{23700}. The m-CU enables seamless transitions between \gls{iab}-donors within their coverage areas, ensuring uninterrupted handovers and maintaining stable \gls{ue} connectivity. 
This innovation enables the \gls{5g} network to effectively support \gls{m-iab}-nodes in highly dynamic environments.


\section{End-to-end simulation of IAB networks}
\label{sec:simulation-module}
In this work, we designed and developed an open-source ns-3 module for \gls{5g} \gls{iab} networks, named \texttt{ns3-mmwave-iab}\footnote{\url{https://github.com/signetlabdei/ns3-IAB.git}.}.
This module is built upon the \texttt{ns3-mmwave} framework for \gls{3gpp} \gls{nr} networks~\cite{mezzavilla2018end} and a previous version of the code presented in~\cite{polese2018end}, that we extended to incorporate the latest \gls{iab} functionalities from Release 16,\footnote{The module is compliant with the latest \gls{3gpp} \gls{iab} specifications from the \gls{ran} perspective.
The \gls{cn} is modeled according to the \gls{lte} specifications as defined in the \texttt{ns3-mmwave} module~\cite{mezzavilla2018end}.} as described in \cref{sec:iab-standard}.

Notably, the module was extended to simulate maritime scenarios, especially through a dedicated maritime-specific channel model and offshore scenario configurations. Still, the module remains sufficiently general to support future studies combining maritime and terrestrial segments and comparisons with alternative architectures (e.g., terrestrial-only \gls{iab}, donor-based deployments, and relays).

In the remainder of this section, we present the main components of the proposed IAB simulation framework, specifically our \gls{bap} adaptation layer implementation (\cref{sub:bap-implementation}), the end-to-end protocol stack and data flow (\cref{sub:stack-implementation}), a maritime-specific channel model (\cref{sec:channel_model}), the slot format and control signaling structure (\cref{sec:slot_format}), and the \gls{mt} and \gls{du} multiplexing schemes (\cref{sec:duplexing}).

\subsection{BAP Implementation}
\label{sub:bap-implementation}
First, we implemented in ns-3 the \gls{bap} adaptation layer (see \cref{sec:3gpp_bap}). This layer is located on top of the \gls{rlc} layer, and is in charge of routing packets between the \gls{iab}-donor and the access \gls{iab}-node via wireless backhauling.

In the module, we model the BAP \emph{Data \glspl{pdu}}, which are used to transport upper-layer data~\cite[Sec.~6.1.1]{38340}, while the implementation of \emph{Control PDUs} is left for future work.
The header of a BAP \emph{Data PDU} is of 3 bytes, and comprises the following fields~\cite[Sec.~6.3]{38340}:
\begin{itemize}
    \item D/C (1 bit), indicating whether the packet represents a BAP \textit{Data PDU} or \textit{Control PDU};
    \item DESTINATION (10 bits), i.e., the BAP address of the destination \gls{iab}-node (DL) or \gls{iab}-donor (UL) \gls{du}; 
    \item PATH (10 bits), representing the \gls{bap} path ID; 
    \item R (3 bits), reserved for future use.
\end{itemize}
Given that both the DESTINATION and the PATH fields are of 10 bits, the system can support up to 1024 potential pathways, and up to 1024 \gls{iab}-nodes, including the \gls{iab}-donor.
Additionally, we implemented the logic for handling \gls{bap} \glspl{pdu} and \glspl{sdu}, as described in \cref{sub:stack-implementation}.

\subsection{Protocol Stack and End-to-End Data Flow}
\label{sub:stack-implementation}

In line with the 3GPP~\cite{171880}, in our simulator we reuse most \gls{5g} \gls{nr} primitives to model the IAB protocol stack. Specifically, we model IAB-nodes as wireless backhauled base stations which feature two interfaces, i.e., the DU and the MT, as depicted in \cref{fig:iab-stack-sim}. The former provides connectivity to the UEs and the child MTs, like traditional base stations, while the latter connects to upstream DUs.

\subsubsection{User Plane (PDU traffic)}
In the \gls{up}, each interface implements the 5G NR \gls{rlc}, \gls{mac}, and \gls{phy} layers. Additionally, both MTs and DUs also implement the BAP adaptation layer, to handle routing and forwarding. Finally, we introduced in both IAB-node and IAB-donor DUs a so-called ``inner \gls{pdcp}'' entity, as illustrated in \cref{fig:iab-stack-sim-pdu}, which represents the \gls{gtp}-U tunnel endpoint for the backhaul network.

Accordingly, in our ns-3 implementation, \gls{dl} \gls{up} packets are handled as follows.

\begin{enumerate}
    \item Packets enter the \gls{cn} from the \gls{dn} via the \gls{pgw} and \gls{sgw}, which together emulate the functionality of the \gls{5gc} \gls{upf}. 
   \item Packets are then encapsulated in GTP-U tunnels.
    Each tunnel is identified by a \gls{teid} associated with a data bearer.
    In our implementation, \glspl{tft}~\cite{24008} are used to classify packets into the corresponding tunnels according to LTE S1-U interface procedures~\cite{piro2011lte}. 
    Therefore, the role of \glspl{tft} is similar to that of the \gls{sdap} layer in 5G NR, i.e., mapping a \gls{qos} flow from the \gls{5gc} to a data bearer via the GTP-U extension header~\cite[Sec.~6.5]{38300}.  
   \item Once \glspl{teid} are mapped to radio bearers, packets are forwarded to the corresponding \gls{pdcp} instance.
   Then, the ``inner \gls{pdcp}'' entity encapsulates \gls{up} packets in internal GTP-U tunnels, associating radio bearers to TEIDs and storing this mapping at the IAB-donor~\cite[Sec.~6.3.1]{3gpp_38_874}.
   GTP-U, \gls{udp}, and IP headers are then appended and packets are encapsulated over the \gls{tnl}; the IP \gls{dscp} field is set according to the assigned TEID.   
   \item Subsequently, the PDCP instance forwards the encapsulated \gls{up} packet to the BAP layer. The latter leverages the \gls{tnl} header to:
(i) determine the BAP destination address and path ID;
    (ii) encode them in a BAP header; and
    (iii) forward the BAP \emph{Data PDU} to the corresponding RLC channel~\cite{3gpp_38_874}~\cite[Sec. 6.3.2]{38340}.
In our implementation, the mapping between PDCP entities and the corresponding \gls{bap} destination address and path ID is handled through a forwarding table maintained by the IAB-donor.
\item Once the \gls{bap} PDU is generated, it is propagated across the IAB backhaul. At each intermediate IAB-node, the local \gls{bap} instance checks whether the destination \gls{bap} address matches its own. If the address does not match, the node identifies the appropriate egress RLC channel associated with the indicated path ID and forwards the PDU to the next hop.
When the \gls{bap} PDU reaches the destination IAB-node, 
the outer \gls{bap}, GTP-U, \gls{udp}, and IP headers are removed, thus terminating the internal GTP-U tunnel.
\item The packet is then passed to the \gls{rlc} entity of the serving DU and transmitted over the access link to the destination~UE.
\end{enumerate}

\gls{ul} \gls{up} packets are handled as in the \gls{dl} case, with the only difference that the invocation of the ``inner \gls{pdcp}'' instance, and the corresponding encapsulation into the internal GTP-U tunnel, are performed at the access IAB-node rather than at the IAB-donor.



\begin{figure*}[t!]
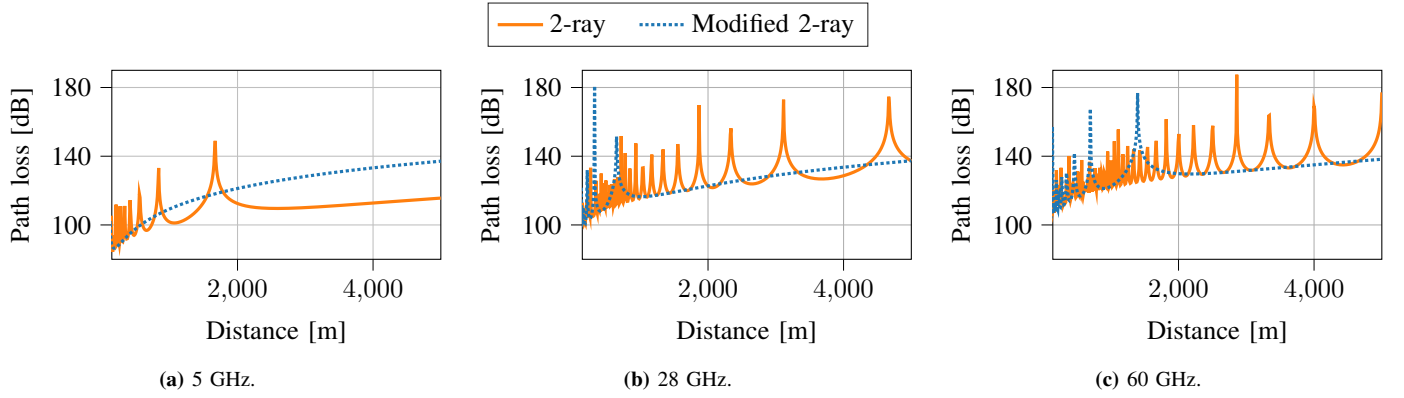

    \begin{subfigure}[b]{\linewidth}
	\centering
	\setlength\fwidth{\columnwidth}
%
%

\definecolor{plotColor1}{RGB}{255,127,14}
\definecolor{plotColor2}{RGB}{31,119,180}

\begin{tikzpicture}

\pgfplotsset{every tick label/.append style={font=\scriptsize}}

\pgfplotsset{compat=1.11,
	/pgfplots/ybar legend/.style={
		/pgfplots/legend image code/.code={%
			\draw[##1,/tikz/.cd,yshift=-0.25em]
			(0cm,0cm) rectangle (10pt,0.6em);},
	},
}

\begin{axis}[%
width=0,
height=0,
at={(0,0)},
scale only axis,
xtick={},
ytick={},
axis background/.style={fill=white},
legend style={legend cell align=left,
              align=center,
              draw=white!15!black,
              at={(0.5, 1.3)},
              anchor=center,
              /tikz/every even column/.append style={column sep=1em}},
legend columns=2,
]
\addplot[plotColor1, very thick]
table[row sep=crcr]{%
	0	0\\
};
\addlegendentry{2-ray}

\addplot[plotColor2, very thick, densely dotted]
table[row sep=crcr]{%
	0	0\\
};
\addlegendentry{Modified 2-ray}

\end{axis}
\end{tikzpicture}%
     \end{subfigure}
    \vskip 0.3cm
    \centering
    \subfloat[][5 GHz.]
	{
        \begin{minipage}{0.3\textwidth}
\begin{tikzpicture}

\definecolor{darkgray176}{RGB}{176,176,176}
\definecolor{plotColor1}{RGB}{255,127,14}
\definecolor{plotColor2}{RGB}{31,119,180}

\begin{axis}[
  height = 2.5cm,
  width=0.8\linewidth,
  tick pos=left,
  tick align=outside,
  scale only axis,
  xlabel={Distance [m]},
  xmajorgrids,
  xmin=145, xmax=5000,
  xtick style={color=black},
  ylabel={Path loss [dB]},
  ymajorgrids,
  ymin=80, ymax=190,
  ytick style={color=black},
  ytick={100, 140, 180},
  yticklabels={100, 140, 180}
]
\addplot [plotColor1, very thick]
table {%
100 81.8454666137695
109.819641113281 86.3497314453125
119.639282226562 96.568244934082
129.458923339844 91.5396575927734
139.278564453125 105.291793823242
149.098190307617 89.0405960083008
158.917831420898 84.4311676025391
168.73747253418 93.9199447631836
178.557113647461 86.5808715820312
188.376754760742 92.9640808105469
198.196395874023 86.6620178222656
208.016036987305 111.846206665039
217.835678100586 88.2616500854492
227.655303955078 88.9188690185547
237.474945068359 110.695289611816
247.294586181641 91.1274948120117
257.114227294922 88.61669921875
266.933868408203 91.998046875
276.753509521484 110.891471862793
286.573150634766 94.9655227661133
296.392791748047 90.5551452636719
306.212432861328 90.2497863769531
316.032073974609 92.7253036499023
325.851715087891 99.4622344970703
335.671356201172 111.060523986816
345.490966796875 96.9210433959961
355.310607910156 93.1425704956055
365.130249023438 91.8537368774414
374.949890136719 92.001953125
384.76953125 93.3401870727539
394.589172363281 96.0237426757812
404.408813476562 100.946044921875
414.228454589844 114.40926361084
424.048095703125 106.566474914551
433.867736816406 99.7026519775391
443.687377929688 96.6110687255859
453.507019042969 94.9669418334961
463.32666015625 94.1572341918945
473.146301269531 93.933219909668
482.965942382812 94.1787796020508
492.785583496094 94.8448104858398
502.605224609375 95.9296188354492
512.424865722656 97.4814910888672
522.244506835938 99.6267013549805
532.064147949219 102.660980224609
541.8837890625 107.406631469727
551.703430175781 118.056312561035
561.523071289062 115.971405029297
571.342712402344 107.538841247559
581.162353515625 103.700973510742
590.98193359375 101.349296569824
600.801574707031 99.7621154785156
610.621215820312 98.6521301269531
620.440856933594 97.8750228881836
630.260498046875 97.3478469848633
640.080139160156 97.0186920166016
649.899780273438 96.8531494140625
659.719421386719 96.8275985717773
669.5390625 96.9255752563477
679.358703613281 97.1356811523438
689.178344726562 97.4503860473633
698.997985839844 97.8653182983398
708.817626953125 98.3789520263672
718.637268066406 98.9925689697266
728.456909179688 99.7104797363281
738.276550292969 100.540618896484
748.09619140625 101.495597839355
757.915832519531 102.594566345215
767.735473632812 103.866439819336
777.555114746094 105.355880737305
787.374755859375 107.135116577148
797.194396972656 109.330055236816
807.014038085938 112.187980651855
816.833679199219 116.305313110352
826.6533203125 123.925270080566
836.472961425781 133.167617797852
846.292602539062 119.699989318848
856.112243652344 114.854850769043
865.931884765625 111.911758422852
875.751525878906 109.836975097656
885.571166992188 108.264053344727
895.390808105469 107.019874572754
905.21044921875 106.008666992188
915.030090332031 105.171768188477
924.849670410156 104.470558166504
934.669311523438 103.87825012207
954.30859375 102.947654724121
973.947875976562 102.274955749512
993.587158203125 101.796096801758
1013.22644042969 101.46900177002
1032.86572265625 101.264533996582
1062.32470703125 101.14330291748
1091.78356933594 101.202033996582
1121.24243164062 101.407302856445
1160.52099609375 101.870178222656
1199.79956054688 102.51725769043
1239.078125 103.327178955078
1278.35668945312 104.289505004883
1317.63525390625 105.403160095215
1356.91381835938 106.676574707031
1396.1923828125 108.129585266113
1425.65124511719 109.357986450195
1455.11022949219 110.731140136719
1484.56909179688 112.285461425781
1504.20837402344 113.449295043945
1523.84765625 114.743553161621
1543.48693847656 116.204849243164
1563.12622070312 117.889083862305
1572.94592285156 118.841178894043
1582.76550292969 119.887557983398
1592.58520507812 121.051170349121
1602.40478515625 122.36466217041
1612.22448730469 123.876724243164
1622.04406738281 125.664642333984
1631.86376953125 127.862274169922
1641.68334960938 130.733627319336
1651.50305175781 134.927093505859
1661.32263183594 143.026596069336
1671.14233398438 148.956680297852
1680.9619140625 137.113906860352
1690.78161621094 132.369873046875
1700.60119628906 129.380630493164
1710.4208984375 127.208305358887
1720.24047851562 125.510482788086
1730.06018066406 124.123191833496
1739.87976074219 122.955078125
1749.69934082031 121.949989318848
1759.51904296875 121.070938110352
1769.33862304688 120.292236328125
1788.97790527344 118.966415405273
1808.6171875 117.872360229492
1828.25646972656 116.949340820312
1847.89575195312 116.157707214355
1877.35473632812 115.159057617188
1906.81359863281 114.333824157715
1936.27258300781 113.641311645508
1975.55114746094 112.87718963623
2014.82971191406 112.254318237305
2063.927734375 111.627891540527
2122.845703125 111.046272277832
2191.58325195312 110.543167114258
2270.14038085938 110.13932800293
2358.51708984375 109.845306396484
2466.53295898438 109.652824401855
2594.1884765625 109.593063354492
2751.30249023438 109.689956665039
2937.87573242188 109.965179443359
3183.36669921875 110.488861083984
3527.05419921875 111.386421203613
4194.78955078125 113.31827545166
5000 115.631736755371
};
\addplot [plotColor2, very thick, densely dotted, forget plot]
table {%
100 86.8866806030273
109.819641113281 96.5279159545898
119.639282226562 102.236328125
129.458923339844 92.5335388183594
139.278564453125 89.3078460693359
149.098190307617 87.7488708496094
158.917831420898 86.9503479003906
168.73747253418 86.5740203857422
178.557113647461 86.4568252563477
188.376754760742 86.5083923339844
198.196395874023 86.6740188598633
217.835678100586 87.2184066772461
247.294586181641 88.3084716796875
316.032073974609 91.1676635742188
374.949890136719 93.5271377563477
424.048095703125 95.3453369140625
473.146301269531 97.0261001586914
532.064147949219 98.8766632080078
590.98193359375 100.567306518555
659.719421386719 102.366287231445
728.456909179688 104.006317138672
807.014038085938 105.716567993164
885.571166992188 107.27906036377
973.947875976562 108.888191223145
1072.14428710938 110.520790100098
1180.16027832031 112.158569335938
1297.99597167969 113.787551879883
1425.65124511719 115.397331237793
1563.12622070312 116.980422973633
1720.24047851562 118.630317687988
1887.17431640625 120.228103637695
2073.74755859375 121.856460571289
2279.9599609375 123.495498657227
2505.8115234375 125.129905700684
2751.30249023438 126.74821472168
3016.43286132812 128.34211730957
3311.02197265625 129.957214355469
3635.07006835938 131.576217651367
3988.5771484375 133.185943603516
4381.36279296875 134.815490722656
4803.607421875 136.412124633789
5000 137.107574462891
};
\end{axis}

\end{tikzpicture}
        \end{minipage}
        \vspace{-0.4cm}
	}
    \hfill
   \subfloat[][28 GHz.]
	{
        \begin{minipage}{0.3\textwidth}
            \input{Figs/PL-28.0-GHz}
        \end{minipage}
        \vspace{-0.4cm}
	}
    \hfill
    \subfloat[][60 GHz.]
	{
        \begin{minipage}{0.3\textwidth}
            \input{Figs/PL-60.0-GHz}
        \end{minipage}
        \vspace{-0.4cm}
	}
    \caption{Comparison of the path loss obtained using the 2-ray model and using the modified model proposed in~\cite{mehrnia2016novel}, at different frequencies.}
   \label{fig:pl-comparison}
   \vskip -0.1em
\end{figure*}

\subsubsection{User Plane (non-PDU traffic)}
In addition to DL and UL traffic generated at the \gls{dn} (in the \gls{5gc} or the UE, respectively), our simulator can support traffic that is not attached to a UPF, but rather terminates at the IAB-donor DU, as represented in \cref{fig:iab-stack-sim-lan} (in IAB-node 3).
Such traffic may originate from, or be destined to, physical \gls{lan} interfaces located onboard IAB-nodes.
To enable this functionality, we extend our baseline BAP implementation by using one of the three reserved (R) bits in the BAP header (see \cref{sub:bap-implementation}) to distinguish between PDU and non-PDU traffic.
In the latter case, routing within the backhaul is handled as follows.
When a data packet is received at the donor CU-UP, it is forwarded directly to the outbound BAP interface, bypassing the SDAP, PDCP, and ``inner PDCP'' layers at the IAB-donor. 
The packet is then transported through the IAB backhaul according to a pre-configured BAP destination address and path ID, and finally delivered to the \gls{lan} interface at the target IAB-node.


\subsubsection{Control Plane}

To simplify the \gls{cp} design, we introduce an ideal logical link between each IAB-node DU and the \gls{amf}.
This approach eliminates the need to transmit \gls{cp} signals across the multi-hop \gls{iab} backhaul, which would introduce additional complexity. As such, the SRB0 and SRB1 radio bearers, which carry \gls{cp} data, are directly terminated at the access IAB-node DU~\cite{3gpp_38_874}.
While assuming an ideal logical link does not allow us to capture the delay introduced by multi-hop control-plane signaling and handover procedures, maritime scenarios are generally characterized by low mobility and relatively stable topologies, which supports our assumption. Nevertheless, control signaling overhead is still captured through configurable control symbols in the \gls{nr} frame structure.
Moreover, notice that, while the \gls{amf} is a \gls{cp} function, this abstraction permits focusing on the \gls{up} performance of the IAB network, which remains the primary objective of this work.


\subsection{Maritime Channel Model}
\label{sec:channel_model}

The 3GPP 38.901 channel model~\cite{3gpp.38.901}, which is the de facto standard for system-level simulations of 5G and beyond systems, does not explicitly model maritime propagation scenarios.
Rather, it supports five target environments, namely \gls{uma}, \gls{umi}, \gls{rma}, \gls{inf}, and indoor-office, primarily for terrestrial deployments.

Among these, we identified the \gls{rma} scenario as the closest approximation to maritime conditions, due to the absence of large-scale obstructions and the reduced presence of scatterers.
However, \gls{rma} does not fully capture the effects introduced by the sea surface (e.g., two-ray reflections and path loss variations with antenna height and sea state)~\cite{brata2021path}.
Therefore, we extended the ns-3 implementation of the 3GPP TR~38.901 model in~\cite{zugno20implementation} with the following modifications to better represent coastal and offshore radio environments.

\subsubsection{Path loss and propagation models}


Electromagnetic propagation in maritime and coastal environments is characterized by strong first-order reflections from the sea surface\footnote{In most practical scenarios, the magnitude and phase of the reflection coefficient for vertically polarized signals are approximately $1$ and $180^\circ$, respectively~\cite{reyes2011buoy}.}, as well as by the formation of evaporation ducts.
Accordingly, the 2-ray model~\cite{reyes2011buoy} initially appeared as the most accurate theoretical model to represent the maritime scenario, as demonstrated through experimental results at $5$~GHz~\cite{yee2014near, reyes2011buoy}. 
However, subsequent studies have shown that the 2-ray model becomes inaccurate at higher frequencies, such as in the \gls{5g} \gls{nr} \gls{fr2} bands (e.g., n257 and n263)~\cite{mehrnia2016novel}.
Therefore, in this work we employ the modified 2-ray model proposed in~\cite{mehrnia2016novel}, which characterizes the path loss $PL$ as~\cite[Eq.~5]{mehrnia2016novel}
\begin{equation}
    PL\, \mathrm{[dB]} = -20 \log_{10} \left\{ \frac{\lambda}{4 \pi d} \left\vert 1 + R \exp \left( j\frac{\alpha 2\pi \Delta d}{\lambda} \right) \right\rvert  \right\},
\end{equation}
where $\lambda$ is the wavelength, $d$ is the 3D distance between the transmitter and the receiver, and $\Delta d$ is the difference in length between the direct and first-order reflection paths. $R$ and $\alpha$ are model parameters, which represent the reflection and unit-less frequency-dependent coefficients, respectively. In~\cite{mehrnia2016novel}, $R$ is assumed to be $-1$, while $\alpha$ is approximated via least-square minimization as $\alpha = 1.091 \exp(-0.06256 f) + 0.06982$, where $f$ is the carrier frequency in GHz. As depicted in \cref{fig:pl-comparison}, the modified 2-ray model leads to fewer path loss peaks for long ($> 2$~km) communication links than the classical 2-ray model~\cite{mehrnia2016novel}.
Finally, in accordance with the study in~\cite{mehrnia2016novel}, we consider deterministic \gls{los} conditions, which is a common assumption for coastal and maritime areas.

Although we focus on the maritime scenario, we did not implement the evaporation duct effect. In fact, this effect is primarily experienced for over-the-horizon propagation~\cite{Woods09Duct}, while for short-range (up to 4~km) mostly-\gls{los} links, such as those considered in this work, the duct contribution is negligible compared to other propagation phenomena, such as sea surface reflections and rain attenuation.

\subsubsection{Rain attenuation model}
\label{sec:rain}
In general, coastal areas are affected by more frequent precipitation than inland and oceanic areas, thus experiencing higher average rain rates~\cite{ogino2016much}. In order to capture this peculiarity, we extended our simulator to account for rain-induced attenuation.
Specifically, we implemented the ITU-R P.838-3 model~\cite{itu.p.838}, which characterizes the signal loss $\gamma_{r}$ due to precipitation as~\cite{itu.p.838}
\begin{equation}
\gamma_{r} \, \mathrm{[dB/km]} = k\rho^{\alpha},
\end{equation}
where $\rho$ is the rain rate in mm/h. $k$ and $\alpha$ are model parameters which can be computed, respectively, as
\begin{align}
 \log_{10}k =& \sum_{j=1}^{4} \left( a_j \exp \left[- \left( \frac{\log_{10} f - b_j}{c_j} \right)^2 \right] \right) \nonumber \\
 &+ m_k \log_{10} f + c_k,
\end{align}
and
\begin{align}
 \alpha =& \sum_{j=1}^{5} \left( a_j \exp \left[ - \left( \frac{\log_{10} f - b_j}{c_j} \right)^2 \right] \right) \nonumber \\
 &+ m_{\alpha} \log_{10} f + c_{\alpha},
\end{align}
where $f$ is the carrier frequency.
The coefficients 
$\{a_j, b_j, c_j\}_{j=1,\ldots,5}$, $m_k, c_k$, and $m_{\alpha}, c_{\alpha}$ 
depend on the signal polarization (vertical or horizontal) and are reported in
\cite[Tables~1–4]{itu.p.838}.

To estimate $\rho$, the ITU provides historical rainfall data~\cite{itu.p.837} which can be used to obtain a map of the average rain rates (mm/h) for each location on Earth. 
However, this map accounts for both rainy and non-rainy periods, resulting in an average loss of only 0.2 dB/km, while severe storms can cause up to 15 dB/km of attenuation. Therefore, we treat $\rho$ as a simulation parameter, which can be set at the beginning of the simulation based on the specific scenario and weather conditions.

\subsection{Slot Format and Control Symbols}
\label{sec:slot_format}
In our simulator, we define three slot types based on the traffic direction: 
(i) DL slots for \gls{dl} transmissions; (ii) UL slots for \gls{ul} transmissions; and (iii) switching (SW) slots in between \gls{dl} and \gls{ul} slots.
In an SW slot, the last 4 \gls{ofdm} symbols are allocated for \gls{ul} transmissions, while the preceding symbols act as a guard period to accommodate propagation delays and switching times.
This configuration reflects realistic \gls{tdd} operations, even though the slot format remains fully configurable in the simulator.

Based on the assumptions in the \texttt{ns3-mmwave} baseline module~\cite{mezzavilla2018end}, we reserve the first and last symbols of each slot for control signals, while the rest is available for data transmissions.
However, to enhance the accuracy and flexibility of the simulations, we introduced an option to allocate additional control symbols (\gls{dl} or \gls{ul}) at a custom periodicity and equally distributed across all the slots.

\subsection{MT and DU Multiplexing}
\label{sec:duplexing}
\subsubsection{Time Division Multiplexing}
To reduce interference, our simulator assumes that each \gls{iab}-node can activate only one interface at a time.\footnote{The \gls{3gpp} TS 38.300 specifications define IAB-nodes as half-duplex devices~\cite{38300}, while the \gls{3gpp} TS 38.340 specifications assume that, for in-band IAB, IAB-MT and IAB-DU transmissions and receptions are mutually exclusive in time~\cite{38340}.}
As a result, the \gls{mt} and \gls{du} resources of the same \gls{iab}-node must be orthogonal (in either time or frequency).
Let ``layer'' refer to the depth of an \gls{iab}-node in the network topology, with the \gls{iab}-donor at the root (layer 0). In our setup, the resources allocated to an \gls{iab}-node of an even layer must be orthogonal to those allocated to an \gls{iab}-node of an odd layer.
To implement this system, we assign $n^{o}_{s}$ symbols to \gls{iab}-nodes of the odd layers, while the remaining are allocated to even layers.
For example, using \gls{tdm}, if there is a total of 12~\gls{ofdm} symbols for data transmission, the first $n^{o}_{s}$ symbols can be scheduled to \gls{iab}-nodes of an odd layer, while the rest can be used by \gls{iab}-nodes of an even layer.

\subsubsection{Frequency Division Multiplexing}
In addition to \gls{tdm}, where all transmissions share the same spectrum resources, and interference is mitigated by the fact that the \gls{mt} and \gls{du} of each \gls{iab}-node operate on orthogonal \gls{ofdm} symbols, we implemented \gls{fdm}. 
Specifically, the \gls{mt} and \gls{du} of each \gls{iab}-node can operate simultaneously, albeit in orthogonal frequency bands, through carrier aggregation. 
This approach allows a flexible bandwidth division, supporting both intra- and inter-band aggregation.
Consequently, the \gls{du} of each \gls{iab}-node can schedule both \gls{dl} and \gls{ul} transmissions over the entire slot, using all of the available \gls{ofdm} symbols (i.e., 12 out of 14, since 2 ~\gls{ofdm} symbols are used for control signals, as described in \cref{sec:slot_format}).

\eject

\begin{table}[t]
\caption{Simulation parameters.}
\label{tab:parameters}
\centering
\renewcommand{\arraystretch}{1.3}
\footnotesize
\begin{tabular}{p{0.45\columnwidth} c}
\hline
\textbf{Parameter} & \textbf{Value} \\ 
\hline
DL source rate ($r_{\rm DL}$) [Mb/s] & \{60, 80, 100, 120, 140\} \\
Inter-packet interval [$\mu$s] & 50 \\
Carrier frequency ($f$) [GHz] & 26 \\
Transmit power [dBm] & 30 \\
Bandwidth [MHz] & 400 \\
Rain rate ($\rho$) [mm/h] & \{0, 15, 30\} \\
Beamforming technique & Codebook-based analog \\
IAB-node velocity [m/s] & 5 \\
Simulation time [s] & 2 \\
Simulation runs & 50 \\
Antenna radiation pattern & 3GPP model~\cite{3gpp.38.901}, 13 dBi gain \\
Channel model & Maritime model (see \cref{sec:channel_model}) \\
5G NR numerology index & 3 \\
TDD slot pattern & 4DS2U \\
Symbols for odd-layer nodes ($n^{o}_{s}$) & 6 \\
\hline
\end{tabular}
\vskip -0.1em
\end{table}

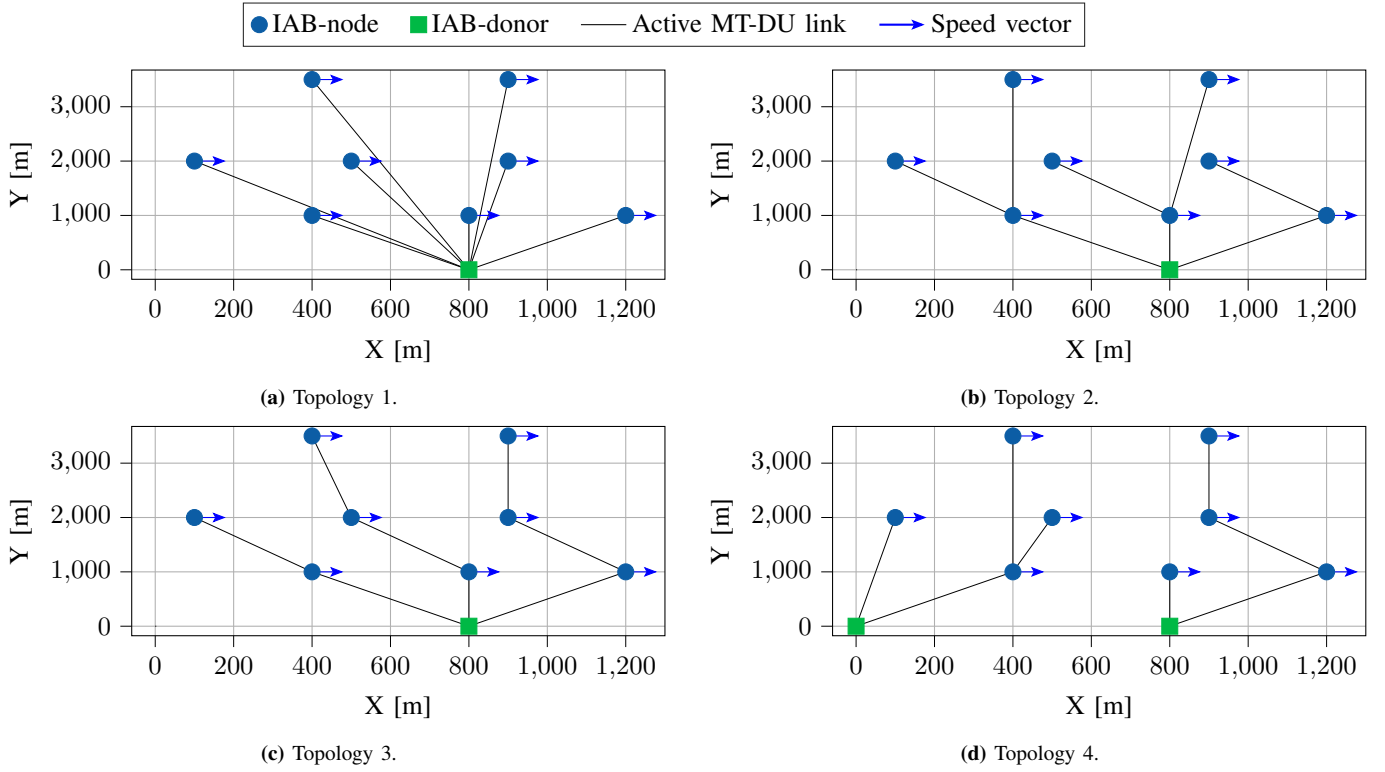
\begin{figure*}[t!]
    \begin{subfigure}[b]{\linewidth}
	\centering
	\setlength\fwidth{\columnwidth}
%
%

\definecolor{color1}{RGB}{105,153,196}
\definecolor{color2}{RGB}{114,160,83}
\definecolor{color3}{RGB}{186,97,26}
\definecolor{teal1293165}{RGB}{12,93,165}
\definecolor{limegreen018569}{RGB}{0,185,69}

\begin{tikzpicture}
\pgfplotsset{every tick label/.append style={font=\scriptsize}}

\pgfplotsset{compat=1.11,
	/pgfplots/ybar legend/.style={
		/pgfplots/legend image code/.code={%
			\draw[##1,/tikz/.cd,yshift=-0.25em]
			(0cm,0cm) rectangle (10pt,0.6em);},
	},
}

\begin{axis}[%
width=0,
height=0,
at={(0,0)},
scale only axis,
xtick={},
ytick={},
axis background/.style={fill=white},
legend style={legend cell align=left,
              align=center,
              draw=white!15!black,
              at={(0.5, 1.3)},
              anchor=center,
              /tikz/every even column/.append style={column sep=1em}},
legend columns=4,
]
\addplot [teal1293165, mark=*, mark size=3, mark options={solid}, only marks]
table[row sep=crcr]{%
	0	0\\
};
\addlegendentry{IAB-node}

\addplot [limegreen018569, mark=square*, mark size=3, mark options={solid}, only marks]
table[row sep=crcr]{%
	0	0\\
};
\addlegendentry{IAB-donor}

\addplot [black]
table[row sep=crcr]{%
	0	0\\
};
\addlegendentry{Active MT-DU link}

\addplot[->, blue, thick, arrows={-Stealth[length=5pt, inset=1pt]}] 
table[row sep=crcr]{%
	0 0\\
	1 1\\
};
\addlegendentry{Speed vector}

\end{axis}
\end{tikzpicture}%
    \end{subfigure}
    \vskip 0.3cm
    \centering
    \subfloat[][Topology 1.]
	{
	    \label{fig:scen_1}
\begin{tikzpicture}

\definecolor{darkgray176}{RGB}{176,176,176}
\definecolor{lightgray204}{RGB}{204,204,204}
\definecolor{limegreen018569}{RGB}{0,185,69}
\definecolor{teal1293165}{RGB}{12,93,165}

\begin{axis}[
width = \textwidth/2.1,
height = 4.35cm,
tick pos=left,
tick align=outside,
legend cell align={left},
legend style={
  fill opacity=0.8,
  draw opacity=1,
  text opacity=1,
  at={(0.03,0.03)},
  anchor=south west,
  draw=lightgray204
},
x grid style={darkgray176},
xlabel={X [m]},
xmajorgrids,
xmin=-60, xmax=1300,
xtick style={color=black},
y grid style={darkgray176},
ylabel={Y [m]},
ymajorgrids,
ymin=-175, ymax=3675,
ytick style={color=black}
]
\path [draw=black, fill=black, line width=0pt]
(axis cs:640,400)
--cycle;
\path [draw=black, fill=black, line width=0pt]
(axis cs:800,400)
--cycle;
\path [draw=black, fill=black, line width=0pt]
(axis cs:960,400)
--cycle;
\path [draw=black, fill=black, line width=0pt]
(axis cs:520,800)
--cycle;
\path [draw=black, fill=black, line width=0pt]
(axis cs:680,800)
--cycle;
\path [draw=black, fill=black, line width=0pt]
(axis cs:840,800)
--cycle;
\path [draw=black, fill=black, line width=0pt]
(axis cs:640,1400)
--cycle;
\path [draw=black, fill=black, line width=0pt]
(axis cs:840,1400)
--cycle;
\addplot [line width=0.32pt, black, forget plot]
table {%
400 1000
800 0
};
\addplot [line width=0.32pt, black, forget plot]
table {%
800 1000
800 0
};
\addplot [line width=0.32pt, black, forget plot]
table {%
1200 1000
800 0
};
\addplot [line width=0.32pt, black, forget plot]
table {%
100 2000
800 0
};
\addplot [line width=0.32pt, black, forget plot]
table {%
500 2000
800 0
};
\addplot [line width=0.32pt, black, forget plot]
table {%
900 2000
800 0
};
\addplot [line width=0.32pt, black, forget plot]
table {%
400 3500
800 0
};
\addplot [line width=0.32pt, black, forget plot]
table {%
900 3500
800 0
};
\addplot [black]
table {%
0 0
0 0
};
\addplot [teal1293165, mark=*, mark size=3, mark options={solid}, only marks]
table {%
400 1000
800 1000
1200 1000
100 2000
500 2000
900 2000
400 3500
900 3500
};

\draw [arrows = {-Stealth[length=5pt, inset=1pt]}, blue] (400,1000) -- (480,1000);
\draw [arrows = {-Stealth[length=5pt, inset=1pt]}, blue] (800,1000) -- (880,1000);
\draw [arrows = {-Stealth[length=5pt, inset=1pt]}, blue] (1200,1000) -- (1280,1000);
\draw [arrows = {-Stealth[length=5pt, inset=1pt]}, blue] (100,2000) -- (180,2000);
\draw [arrows = {-Stealth[length=5pt, inset=1pt]}, blue] (500,2000) -- (580,2000);
\draw [arrows = {-Stealth[length=5pt, inset=1pt]}, blue] (900,2000) -- (980,2000);
\draw [arrows = {-Stealth[length=5pt, inset=1pt]}, blue] (400,3500) -- (480,3500);
\draw [arrows = {-Stealth[length=5pt, inset=1pt]}, blue] (900,3500) -- (980,3500);

\addplot [limegreen018569, mark=square*, mark size=3, mark options={solid}, only marks]
table {%
800 0
};
\end{axis}

\end{tikzpicture}
	}
   \subfloat[][Topology 2.]
	{
            \label{fig:scen_2}
\begin{tikzpicture}

\definecolor{darkgray176}{RGB}{176,176,176}
\definecolor{lightgray204}{RGB}{204,204,204}
\definecolor{limegreen018569}{RGB}{0,185,69}
\definecolor{teal1293165}{RGB}{12,93,165}

\begin{axis}[
width = \textwidth/2.1,
height = 4.35cm,
tick pos=left,
tick align=outside,
legend cell align={left},
legend style={
  fill opacity=0.8,
  draw opacity=1,
  text opacity=1,
  at={(0.03,0.03)},
  anchor=south west,
  draw=lightgray204
},
x grid style={darkgray176},
xlabel={X [m]},
xmajorgrids,
xmin=-60, xmax=1300,
xtick style={color=black},
y grid style={darkgray176},
ylabel={Y [m]},
ymajorgrids,
ymin=-175, ymax=3675,
ytick style={color=black}
]
\path [draw=black, fill=black, line width=0pt]
(axis cs:640,400)
--cycle;
\path [draw=black, fill=black, line width=0pt]
(axis cs:800,400)
--cycle;
\path [draw=black, fill=black, line width=0pt]
(axis cs:960,400)
--cycle;
\path [draw=black, fill=black, line width=0pt]
(axis cs:280,1400)
--cycle;
\path [draw=black, fill=black, line width=0pt]
(axis cs:680,1400)
--cycle;
\path [draw=black, fill=black, line width=0pt]
(axis cs:1080,1400)
--cycle;
\path [draw=black, fill=black, line width=0pt]
(axis cs:400,2000)
--cycle;
\path [draw=black, fill=black, line width=0pt]
(axis cs:840,2000)
--cycle;
\addplot [line width=0.32pt, black, forget plot]
table {%
400 1000
800 0
};
\addplot [line width=0.32pt, black, forget plot]
table {%
800 1000
800 0
};
\addplot [line width=0.32pt, black, forget plot]
table {%
1200 1000
800 0
};
\addplot [line width=0.32pt, black, forget plot]
table {%
100 2000
400 1000
};
\addplot [line width=0.32pt, black, forget plot]
table {%
500 2000
800 1000
};
\addplot [line width=0.32pt, black, forget plot]
table {%
900 2000
1200 1000
};
\addplot [line width=0.32pt, black, forget plot]
table {%
400 3500
400 1000
};
\addplot [line width=0.32pt, black, forget plot]
table {%
900 3500
800 1000
};
\addplot [black]
table {%
0 0
0 0
};
\addplot [teal1293165, mark=*, mark size=3, mark options={solid}, only marks]
table {%
400 1000
800 1000
1200 1000
100 2000
500 2000
900 2000
400 3500
900 3500
};

\draw [arrows = {-Stealth[length=5pt, inset=1pt]}, blue] (400,1000) -- (480,1000);
\draw [arrows = {-Stealth[length=5pt, inset=1pt]}, blue] (800,1000) -- (880,1000);
\draw [arrows = {-Stealth[length=5pt, inset=1pt]}, blue] (1200,1000) -- (1280,1000);
\draw [arrows = {-Stealth[length=5pt, inset=1pt]}, blue] (100,2000) -- (180,2000);
\draw [arrows = {-Stealth[length=5pt, inset=1pt]}, blue] (500,2000) -- (580,2000);
\draw [arrows = {-Stealth[length=5pt, inset=1pt]}, blue] (900,2000) -- (980,2000);
\draw [arrows = {-Stealth[length=5pt, inset=1pt]}, blue] (400,3500) -- (480,3500);
\draw [arrows = {-Stealth[length=5pt, inset=1pt]}, blue] (900,3500) -- (980,3500);

\addplot [limegreen018569, mark=square*, mark size=3, mark options={solid}, only marks]
table {%
800 0
};
\end{axis}

\end{tikzpicture}
	}
	\vskip 0.2cm
 \subfloat[][Topology 3.]
	{
		\label{fig:scen_3}
\begin{tikzpicture}

\definecolor{darkgray176}{RGB}{176,176,176}
\definecolor{lightgray204}{RGB}{204,204,204}
\definecolor{limegreen018569}{RGB}{0,185,69}
\definecolor{teal1293165}{RGB}{12,93,165}

\begin{axis}[
width = \textwidth/2.1,
height = 4.35cm,
tick pos=left,
tick align=outside,
legend cell align={left},
legend style={
  fill opacity=0.8,
  draw opacity=1,
  text opacity=1,
  at={(0.03,0.03)},
  anchor=south west,
  draw=lightgray204
},
x grid style={darkgray176},
xlabel={X [m]},
xmajorgrids,
xmin=-60, xmax=1300,
xtick style={color=black},
y grid style={darkgray176},
ylabel={Y [m]},
ymajorgrids,
ymin=-175, ymax=3675,
ytick style={color=black}
]
\path [draw=black, fill=black, line width=0pt]
(axis cs:640,400)
--cycle;
\path [draw=black, fill=black, line width=0pt]
(axis cs:800,400)
--cycle;
\path [draw=black, fill=black, line width=0pt]
(axis cs:960,400)
--cycle;
\path [draw=black, fill=black, line width=0pt]
(axis cs:280,1400)
--cycle;
\path [draw=black, fill=black, line width=0pt]
(axis cs:680,1400)
--cycle;
\path [draw=black, fill=black, line width=0pt]
(axis cs:1080,1400)
--cycle;
\path [draw=black, fill=black, line width=0pt]
(axis cs:460,2600)
--cycle;
\path [draw=black, fill=black, line width=0pt]
(axis cs:900,2600)
--cycle;
\addplot [line width=0.32pt, black, forget plot]
table {%
400 1000
800 0
};
\addplot [line width=0.32pt, black, forget plot]
table {%
800 1000
800 0
};
\addplot [line width=0.32pt, black, forget plot]
table {%
1200 1000
800 0
};
\addplot [line width=0.32pt, black, forget plot]
table {%
100 2000
400 1000
};
\addplot [line width=0.32pt, black, forget plot]
table {%
500 2000
800 1000
};
\addplot [line width=0.32pt, black, forget plot]
table {%
900 2000
1200 1000
};
\addplot [line width=0.32pt, black, forget plot]
table {%
400 3500
500 2000
};
\addplot [line width=0.32pt, black, forget plot]
table {%
900 3500
900 2000
};
\addplot [black]
table {%
0 0
0 0
};
\addplot [teal1293165, mark=*, mark size=3, mark options={solid}, only marks]
table {%
400 1000
800 1000
1200 1000
100 2000
500 2000
900 2000
400 3500
900 3500
};

\draw [arrows = {-Stealth[length=5pt, inset=1pt]}, blue] (400,1000) -- (480,1000);
\draw [arrows = {-Stealth[length=5pt, inset=1pt]}, blue] (800,1000) -- (880,1000);
\draw [arrows = {-Stealth[length=5pt, inset=1pt]}, blue] (1200,1000) -- (1280,1000);
\draw [arrows = {-Stealth[length=5pt, inset=1pt]}, blue] (100,2000) -- (180,2000);
\draw [arrows = {-Stealth[length=5pt, inset=1pt]}, blue] (500,2000) -- (580,2000);
\draw [arrows = {-Stealth[length=5pt, inset=1pt]}, blue] (900,2000) -- (980,2000);
\draw [arrows = {-Stealth[length=5pt, inset=1pt]}, blue] (400,3500) -- (480,3500);
\draw [arrows = {-Stealth[length=5pt, inset=1pt]}, blue] (900,3500) -- (980,3500);

\addplot [limegreen018569, mark=square*, mark size=3, mark options={solid}, only marks]
table {%
800 0
};
\end{axis}

\end{tikzpicture}
	}
 \subfloat[][Topology 4.]
        {
		\label{fig:scen_4}
\begin{tikzpicture}

\definecolor{darkgray176}{RGB}{176,176,176}
\definecolor{lightgray204}{RGB}{204,204,204}
\definecolor{limegreen018569}{RGB}{0,185,69}
\definecolor{teal1293165}{RGB}{12,93,165}

\begin{axis}[
width = \textwidth/2.1,
height = 4.35cm,
tick pos=left,
tick align=outside,
legend cell align={left},
legend style={fill opacity=0.8, draw opacity=1, text opacity=1, draw=lightgray204},
x grid style={darkgray176},
xlabel={X [m]},
xmajorgrids,
xmin=-60, xmax=1300,
xtick style={color=black},
y grid style={darkgray176},
ylabel={Y [m]},
ymajorgrids,
ymin=-175, ymax=3675,
ytick style={color=black}
]
\path [draw=black, fill=black, line width=0pt]
(axis cs:160,400)
--cycle;
\path [draw=black, fill=black, line width=0pt]
(axis cs:800,400)
--cycle;
\path [draw=black, fill=black, line width=0pt]
(axis cs:960,400)
--cycle;
\path [draw=black, fill=black, line width=0pt]
(axis cs:40,800)
--cycle;
\path [draw=black, fill=black, line width=0pt]
(axis cs:440,1400)
--cycle;
\path [draw=black, fill=black, line width=0pt]
(axis cs:1080,1400)
--cycle;
\path [draw=black, fill=black, line width=0pt]
(axis cs:400,2000)
--cycle;
\path [draw=black, fill=black, line width=0pt]
(axis cs:900,2600)
--cycle;
\addplot [line width=0.32pt, black, forget plot]
table {%
400 1000
0 0
};
\addplot [line width=0.32pt, black, forget plot]
table {%
800 1000
800 0
};
\addplot [line width=0.32pt, black, forget plot]
table {%
1200 1000
800 0
};
\addplot [line width=0.32pt, black, forget plot]
table {%
100 2000
0 0
};
\addplot [line width=0.32pt, black, forget plot]
table {%
500 2000
400 1000
};
\addplot [line width=0.32pt, black, forget plot]
table {%
900 2000
1200 1000
};
\addplot [line width=0.32pt, black, forget plot]
table {%
400 3500
400 1000
};
\addplot [line width=0.32pt, black, forget plot]
table {%
900 3500
900 2000
};
\addplot [black]
table {%
0 0
0 0
};
\addplot [teal1293165, mark=*, mark size=3, mark options={solid}, only marks]
table {%
400 1000
800 1000
1200 1000
100 2000
500 2000
900 2000
400 3500
900 3500
};

\draw [arrows = {-Stealth[length=5pt, inset=1pt]}, blue] (400,1000) -- (480,1000);
\draw [arrows = {-Stealth[length=5pt, inset=1pt]}, blue] (800,1000) -- (880,1000);
\draw [arrows = {-Stealth[length=5pt, inset=1pt]}, blue] (1200,1000) -- (1280,1000);
\draw [arrows = {-Stealth[length=5pt, inset=1pt]}, blue] (100,2000) -- (180,2000);
\draw [arrows = {-Stealth[length=5pt, inset=1pt]}, blue] (500,2000) -- (580,2000);
\draw [arrows = {-Stealth[length=5pt, inset=1pt]}, blue] (900,2000) -- (980,2000);
\draw [arrows = {-Stealth[length=5pt, inset=1pt]}, blue] (400,3500) -- (480,3500);
\draw [arrows = {-Stealth[length=5pt, inset=1pt]}, blue] (900,3500) -- (980,3500);

\addplot [limegreen018569, mark=square*, mark size=3, mark options={solid}, only marks]
table {%
800 0
0 0
};
\end{axis}

\end{tikzpicture}
	}
    \vskip 0.2cm
    \caption{IAB network topologies. The arrows represent the speed vectors of \gls{iab}-nodes.}
   \label{fig:scenarios}
   \vskip -0.1em
\end{figure*}

\section[Performance Evaluation of a Maritime IAB Scenario]
{Performance Evaluation of\\ a Maritime IAB Scenario}
\label{sec:perf-eval}

\begin{figure*}[t!]
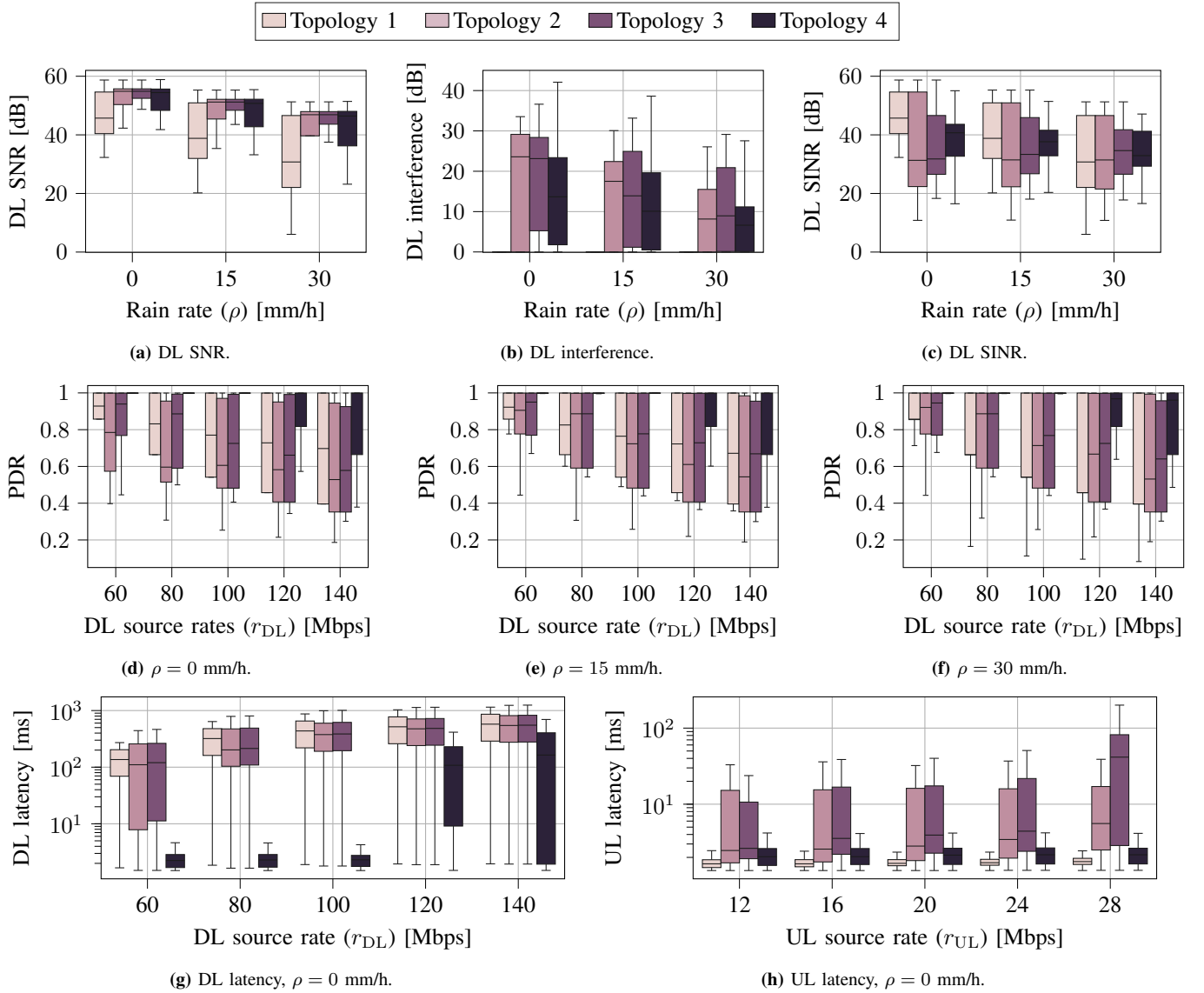

    \begin{subfigure}[b]{\linewidth}
	\centering
	\setlength\fwidth{\columnwidth}
%
%

\definecolor{color4}{RGB}{44,34,57}
\definecolor{color3}{RGB}{124,82,119}
\definecolor{color1}{RGB}{233,211,207}
\definecolor{color2}{RGB}{191,143,160}

\begin{tikzpicture}
\pgfplotsset{every tick label/.append style={font=\scriptsize}}

\pgfplotsset{compat=1.11,
	/pgfplots/ybar legend/.style={
		/pgfplots/legend image code/.code={%
			\draw[##1,/tikz/.cd,yshift=-0.25em]
			(0cm,0cm) rectangle (10pt,0.6em);},
	},
}

\begin{axis}[%
width=0,
height=0,
at={(0,0)},
scale only axis,
xtick={},
ytick={},
axis background/.style={fill=white},
legend style={legend cell align=left,
              align=center,
              draw=white!15!black,
              at={(0.5, 1.3)},
              anchor=center,
              /tikz/every even column/.append style={column sep=1em}},
legend columns=4,
]
\addplot[ybar,ybar legend,draw=black,fill=color1,line width=0.08pt]
table[row sep=crcr]{%
	0	0\\
};
\addlegendentry{Topology 1}

\addplot[ybar legend,ybar,draw=black,fill=color2,fill opacity=0.6,line width=0.08pt]
  table[row sep=crcr]{%
	0	0\\
};
\addlegendentry{Topology 2}

\addplot[ybar legend,ybar,draw=black,fill=color3,line width=0.08pt]
  table[row sep=crcr]{%
	0	0\\
};
\addlegendentry{Topology 3}

\addplot[ybar legend,ybar,draw=black,fill=color4,line width=0.08pt]
  table[row sep=crcr]{%
	0	0\\
};
\addlegendentry{Topology 4}

\end{axis}
\end{tikzpicture}%
    \end{subfigure}
    \vskip 0.2cm
    \centering
   \subfloat[][DL SNR.]
	{
            \label{fig:dl_snr}
\begin{tikzpicture}

\definecolor{black27}{RGB}{27,27,27}
\definecolor{darkgray176}{RGB}{176,176,176}
\definecolor{color4}{RGB}{44,34,57}
\definecolor{color3}{RGB}{124,82,119}
\definecolor{color1}{RGB}{233,211,207}
\definecolor{color2}{RGB}{191,143,160}

\begin{axis}[
width = \textwidth/3.1,
height = 4.35cm,
tick pos=left,
tick align=outside,
x grid style={darkgray176},
xlabel={Rain rate ($\rho$) [mm/h]},
xmajorgrids,
xmin=-0.5, xmax=2.5,
xtick style={color=black},
xtick={0,1,2},
xticklabels={0,15,30},
y grid style={darkgray176},
ylabel={DL SNR [dB]},
ymajorgrids,
ymin=-0.1, ymax=62,
ytick style={color=black},
]

\path [draw=black27, fill=color1]
(axis cs:-0.4,40.416600227356)
--(axis cs:-0.2,40.416600227356)
--(axis cs:-0.2,54.6165990829468)
--(axis cs:-0.4,54.6165990829468)
--(axis cs:-0.4,40.416600227356)
--cycle;
\addplot [black27]
table {%
-0.3 40.416600227356
-0.3 32.293399810791
};
\addplot [black27]
table {%
-0.3 54.6165990829468
-0.3 58.7088012695312
};
\addplot [black27]
table {%
-0.35 32.293399810791
-0.25 32.293399810791
};
\addplot [black27]
table {%
-0.35 58.7088012695312
-0.25 58.7088012695312
};
\path [draw=black27, fill=color1]
(axis cs:0.6,31.9496488571167)
--(axis cs:0.8,31.9496488571167)
--(axis cs:0.8,50.9164991378784)
--(axis cs:0.6,50.9164991378784)
--(axis cs:0.6,31.9496488571167)
--cycle;
\addplot [black27]
table {%
0.7 31.9496488571167
0.7 20.1909999847412
};
\addplot [black27]
table {%
0.7 50.9164991378784
0.7 55.2733993530273
};
\addplot [black27]
table {%
0.65 20.1909999847412
0.75 20.1909999847412
};
\addplot [black27]
table {%
0.65 55.2733993530273
0.75 55.2733993530273
};
\path [draw=black27, fill=color1]
(axis cs:1.6,22.0532252788544)
--(axis cs:1.8,22.0532252788544)
--(axis cs:1.8,46.591700553894)
--(axis cs:1.6,46.591700553894)
--(axis cs:1.6,22.0532252788544)
--cycle;
\addplot [black27]
table {%
1.7 22.0532252788544
1.7 6.04536008834839
};
\addplot [black27]
table {%
1.7 46.591700553894
1.7 51.2579002380371
};
\addplot [black27]
table {%
1.65 6.04536008834839
1.75 6.04536008834839
};
\addplot [black27]
table {%
1.65 51.2579002380371
1.75 51.2579002380371
};
\path [draw=black27, fill=color2]
(axis cs:-0.2,50.3295011520386)
--(axis cs:2.77555756156289e-17,50.3295011520386)
--(axis cs:2.77555756156289e-17,55.6994762420654)
--(axis cs:-0.2,55.6994762420654)
--(axis cs:-0.2,50.3295011520386)
--cycle;
\addplot [black27]
table {%
-0.1 50.3295011520386
-0.1 42.2877006530762
};
\addplot [black27]
table {%
-0.1 55.6994762420654
-0.1 58.7088012695312
};
\addplot [black27]
table {%
-0.15 42.2877006530762
-0.05 42.2877006530762
};
\addplot [black27]
table {%
-0.15 58.7088012695312
-0.05 58.7088012695312
};
\path [draw=black27, fill=color2]
(axis cs:0.8,45.4072513580322)
--(axis cs:1,45.4072513580322)
--(axis cs:1,52.1296491622925)
--(axis cs:0.8,52.1296491622925)
--(axis cs:0.8,45.4072513580322)
--cycle;
\addplot [black27]
table {%
0.9 45.4072513580322
0.9 35.3456993103027
};
\addplot [black27]
table {%
0.9 52.1296491622925
0.9 55.2733993530273
};
\addplot [black27]
table {%
0.85 35.3456993103027
0.95 35.3456993103027
};
\addplot [black27]
table {%
0.85 55.2733993530273
0.95 55.2733993530273
};
\path [draw=black27, fill=color2]
(axis cs:1.8,39.6540484428406)
--(axis cs:2,39.6540484428406)
--(axis cs:2,47.9374237060547)
--(axis cs:1.8,47.9374237060547)
--(axis cs:1.8,39.6540484428406)
--cycle;
\addplot [black27]
table {%
1.9 39.6540484428406
1.9 39.6540484428406
};
\addplot [black27]
table {%
1.9 47.9374237060547
1.9 51.2579002380371
};
\addplot [black27]
table {%
1.85 39.6540484428406
1.95 39.6540484428406
};
\addplot [black27]
table {%
1.85 51.2579002380371
1.95 51.2579002380371
};
\path [draw=black27, fill=color3]
(axis cs:-2.77555756156289e-17,52.4648733139038)
--(axis cs:0.2,52.4648733139038)
--(axis cs:0.2,55.6994762420654)
--(axis cs:-2.77555756156289e-17,55.6994762420654)
--(axis cs:-2.77555756156289e-17,52.4648733139038)
--cycle;
\addplot [black27]
table {%
0.1 52.4648733139038
0.1 48.7112998962402
};
\addplot [black27]
table {%
0.1 55.6994762420654
0.1 58.7088012695312
};
\addplot [black27]
table {%
0.05 48.7112998962402
0.15 48.7112998962402
};
\addplot [black27]
table {%
0.05 58.7088012695312
0.15 58.7088012695312
};
\path [draw=black27, fill=color3]
(axis cs:1,48.379301071167)
--(axis cs:1.2,48.379301071167)
--(axis cs:1.2,52.1296491622925)
--(axis cs:1,52.1296491622925)
--(axis cs:1,48.379301071167)
--cycle;
\addplot [black27]
table {%
1.1 48.379301071167
1.1 43.5466003417969
};
\addplot [black27]
table {%
1.1 52.1296491622925
1.1 55.2733993530273
};
\addplot [black27]
table {%
1.05 43.5466003417969
1.15 43.5466003417969
};
\addplot [black27]
table {%
1.05 55.2733993530273
1.15 55.2733993530273
};
\path [draw=black27, fill=color3]
(axis cs:2,43.6299734115601)
--(axis cs:2.2,43.6299734115601)
--(axis cs:2.2,47.9374237060547)
--(axis cs:2,47.9374237060547)
--(axis cs:2,43.6299734115601)
--cycle;
\addplot [black27]
table {%
2.1 43.6299734115601
2.1 37.5099983215332
};
\addplot [black27]
table {%
2.1 47.9374237060547
2.1 51.2579002380371
};
\addplot [black27]
table {%
2.05 37.5099983215332
2.15 37.5099983215332
};
\addplot [black27]
table {%
2.05 51.2579002380371
2.15 51.2579002380371
};
\path [draw=black27, fill=color4]
(axis cs:0.2,48.3452253341675)
--(axis cs:0.4,48.3452253341675)
--(axis cs:0.4,55.6325492858887)
--(axis cs:0.2,55.6325492858887)
--(axis cs:0.2,48.3452253341675)
--cycle;
\addplot [black27]
table {%
0.3 48.3452253341675
0.3 41.7887001037598
};
\addplot [black27]
table {%
0.3 55.6325492858887
0.3 58.8622016906738
};
\addplot [black27]
table {%
0.25 41.7887001037598
0.35 41.7887001037598
};
\addplot [black27]
table {%
0.25 58.8622016906738
0.35 58.8622016906738
};
\path [draw=black27, fill=color4]
(axis cs:1.2,42.7604236602783)
--(axis cs:1.4,42.7604236602783)
--(axis cs:1.4,52.1527490615845)
--(axis cs:1.2,52.1527490615845)
--(axis cs:1.2,42.7604236602783)
--cycle;
\addplot [black27]
table {%
1.3 42.7604236602783
1.3 33.2000007629395
};
\addplot [black27]
table {%
1.3 52.1527490615845
1.3 55.4267997741699
};
\addplot [black27]
table {%
1.25 33.2000007629395
1.35 33.2000007629395
};
\addplot [black27]
table {%
1.25 55.4267997741699
1.35 55.4267997741699
};
\path [draw=black27, fill=color4]
(axis cs:2.2,36.2328014373779)
--(axis cs:2.4,36.2328014373779)
--(axis cs:2.4,48.0186738967896)
--(axis cs:2.2,48.0186738967896)
--(axis cs:2.2,36.2328014373779)
--cycle;
\addplot [black27]
table {%
2.3 36.2328014373779
2.3 23.1613006591797
};
\addplot [black27]
table {%
2.3 48.0186738967896
2.3 51.4113006591797
};
\addplot [black27]
table {%
2.25 23.1613006591797
2.35 23.1613006591797
};
\addplot [black27]
table {%
2.25 51.4113006591797
2.35 51.4113006591797
};
\addplot [black27]
table {%
-0.4 45.7103500366211
-0.2 45.7103500366211
};
\addplot [black27]
table {%
0.6 38.8129997253418
0.8 38.8129997253418
};
\addplot [black27]
table {%
1.6 30.7453002929688
1.8 30.7453002929688
};
\addplot [black27]
table {%
-0.2 54.865650177002
2.77555756156289e-17 54.865650177002
};
\addplot [black27]
table {%
0.8 51.1919994354248
1 51.1919994354248
};
\addplot [black27]
table {%
1.8 46.8728504180908
2 46.8728504180908
};
\addplot [black27]
table {%
-2.77555756156289e-17 54.865650177002
0.2 54.865650177002
};
\addplot [black27]
table {%
1 51.1919994354248
1.2 51.1919994354248
};
\addplot [black27]
table {%
2 46.8728504180908
2.2 46.8728504180908
};
\addplot [black27]
table {%
0.2 54.418701171875
0.4 54.418701171875
};
\addplot [black27]
table {%
1.2 50.7394504547119
1.4 50.7394504547119
};
\addplot [black27]
table {%
2.2 46.4237003326416
2.4 46.4237003326416
};
\end{axis}

\end{tikzpicture}
        }
    \subfloat[][DL interference.]
	{
	    \label{fig:dl_int}
\begin{tikzpicture}

\definecolor{black27}{RGB}{27,27,27}
\definecolor{darkgray176}{RGB}{176,176,176}
\definecolor{color4}{RGB}{44,34,57}
\definecolor{color3}{RGB}{124,82,119}
\definecolor{color1}{RGB}{233,211,207}
\definecolor{color2}{RGB}{191,143,160}

\begin{axis}[
width = \textwidth/3.1,
height = 4.35cm,
tick pos=left,
tick align=outside,
x grid style={darkgray176},
xlabel={Rain rate ($\rho$) [mm/h]},
xmajorgrids,
xmin=-0.5, xmax=2.5,
xtick style={color=black},
xtick={0,1,2},
xticklabels={0,15,30},
y grid style={darkgray176},
ylabel={DL interference [dB]},
ymajorgrids,
ymin=-0.1, ymax=45,
ytick style={color=black},
ytick={0,10,20,30,40},
]

\path [draw=black27, fill=color1]
(axis cs:-0.4,0)
--(axis cs:-0.2,0)
--(axis cs:-0.2,0)
--(axis cs:-0.4,0)
--(axis cs:-0.4,0)
--cycle;
\addplot [black27]
table {%
-0.3 0
-0.3 0
};
\addplot [black27]
table {%
-0.3 0
-0.3 0
};
\addplot [black27]
table {%
-0.35 0
-0.25 0
};
\addplot [black27]
table {%
-0.35 0
-0.25 0
};
\path [draw=black27, fill=color1]
(axis cs:0.6,0)
--(axis cs:0.8,0)
--(axis cs:0.8,0)
--(axis cs:0.6,0)
--(axis cs:0.6,0)
--cycle;
\addplot [black27]
table {%
0.7 0
0.7 0
};
\addplot [black27]
table {%
0.7 0
0.7 0
};
\addplot [black27]
table {%
0.65 0
0.75 0
};
\addplot [black27]
table {%
0.65 0
0.75 0
};
\path [draw=black27, fill=color1]
(axis cs:1.6,0)
--(axis cs:1.8,0)
--(axis cs:1.8,0)
--(axis cs:1.6,0)
--(axis cs:1.6,0)
--cycle;
\addplot [black27]
table {%
1.7 0
1.7 0
};
\addplot [black27]
table {%
1.7 0
1.7 0
};
\addplot [black27]
table {%
1.65 0
1.75 0
};
\addplot [black27]
table {%
1.65 0
1.75 0
};
\path [draw=black27, fill=color2]
(axis cs:-0.2,0)
--(axis cs:2.77555756156289e-17,0)
--(axis cs:2.77555756156289e-17,29.1479015350342)
--(axis cs:-0.2,29.1479015350342)
--(axis cs:-0.2,0)
--cycle;
\addplot [black27]
table {%
-0.1 0
-0.1 0
};
\addplot [black27]
table {%
-0.1 29.1479015350342
-0.1 33.532299041748
};
\addplot [black27]
table {%
-0.15 0
-0.05 0
};
\addplot [black27]
table {%
-0.15 33.532299041748
-0.05 33.532299041748
};
\path [draw=black27, fill=color2]
(axis cs:0.8,0)
--(axis cs:1,0)
--(axis cs:1,22.4136502742767)
--(axis cs:0.8,22.4136502742767)
--(axis cs:0.8,0)
--cycle;
\addplot [black27]
table {%
0.9 0
0.9 0
};
\addplot [black27]
table {%
0.9 22.4136502742767
0.9 30.0799007415771
};
\addplot [black27]
table {%
0.85 0
0.95 0
};
\addplot [black27]
table {%
0.85 30.0799007415771
0.95 30.0799007415771
};
\path [draw=black27, fill=color2]
(axis cs:1.8,0)
--(axis cs:2,0)
--(axis cs:2,15.5062747001648)
--(axis cs:1.8,15.5062747001648)
--(axis cs:1.8,0)
--cycle;
\addplot [black27]
table {%
1.9 0
1.9 0
};
\addplot [black27]
table {%
1.9 15.5062747001648
1.9 26.0487995147705
};
\addplot [black27]
table {%
1.85 0
1.95 0
};
\addplot [black27]
table {%
1.85 26.0487995147705
1.95 26.0487995147705
};
\path [draw=black27, fill=color3]
(axis cs:-2.77555756156289e-17,5.24460124969482)
--(axis cs:0.2,5.24460124969482)
--(axis cs:0.2,28.378475189209)
--(axis cs:-2.77555756156289e-17,28.378475189209)
--(axis cs:-2.77555756156289e-17,5.24460124969482)
--cycle;
\addplot [black27]
table {%
0.1 5.24460124969482
0.1 0
};
\addplot [black27]
table {%
0.1 28.378475189209
0.1 36.6179008483887
};
\addplot [black27]
table {%
0.05 0
0.15 0
};
\addplot [black27]
table {%
0.05 36.6179008483887
0.15 36.6179008483887
};
\path [draw=black27, fill=color3]
(axis cs:1,1.16429901123047)
--(axis cs:1.2,1.16429901123047)
--(axis cs:1.2,24.925600528717)
--(axis cs:1,24.925600528717)
--(axis cs:1,1.16429901123047)
--cycle;
\addplot [black27]
table {%
1.1 1.16429901123047
1.1 0
};
\addplot [black27]
table {%
1.1 24.925600528717
1.1 33.1665000915527
};
\addplot [black27]
table {%
1.05 0
1.15 0
};
\addplot [black27]
table {%
1.05 33.1665000915527
1.15 33.1665000915527
};
\path [draw=black27, fill=color3]
(axis cs:2,0.116177558898926)
--(axis cs:2.2,0.116177558898926)
--(axis cs:2.2,20.8932509422302)
--(axis cs:2,20.8932509422302)
--(axis cs:2,0.116177558898926)
--cycle;
\addplot [black27]
table {%
2.1 0.116177558898926
2.1 0
};
\addplot [black27]
table {%
2.1 20.8932509422302
2.1 29.1342010498047
};
\addplot [black27]
table {%
2.05 0
2.15 0
};
\addplot [black27]
table {%
2.05 29.1342010498047
2.15 29.1342010498047
};
\path [draw=black27, fill=color4]
(axis cs:0.2,1.77775192260742)
--(axis cs:0.4,1.77775192260742)
--(axis cs:0.4,23.3743495941162)
--(axis cs:0.2,23.3743495941162)
--(axis cs:0.2,1.77775192260742)
--cycle;
\addplot [black27]
table {%
0.3 1.77775192260742
0.3 0
};
\addplot [black27]
table {%
0.3 23.3743495941162
0.3 42.0707988739014
};
\addplot [black27]
table {%
0.25 0
0.35 0
};
\addplot [black27]
table {%
0.25 42.0707988739014
0.35 42.0707988739014
};
\path [draw=black27, fill=color4]
(axis cs:1.2,0.495151519775391)
--(axis cs:1.4,0.495151519775391)
--(axis cs:1.4,19.622875213623)
--(axis cs:1.2,19.622875213623)
--(axis cs:1.2,0.495151519775391)
--cycle;
\addplot [black27]
table {%
1.3 0.495151519775391
1.3 0
};
\addplot [black27]
table {%
1.3 19.622875213623
1.3 38.6144981384277
};
\addplot [black27]
table {%
1.25 0
1.35 0
};
\addplot [black27]
table {%
1.25 38.6144981384277
1.35 38.6144981384277
};
\path [draw=black27, fill=color4]
(axis cs:2.2,0.0995750427246094)
--(axis cs:2.4,0.0995750427246094)
--(axis cs:2.4,11.1716508865356)
--(axis cs:2.2,11.1716508865356)
--(axis cs:2.2,0.0995750427246094)
--cycle;
\addplot [black27]
table {%
2.3 0.0995750427246094
2.3 0
};
\addplot [black27]
table {%
2.3 11.1716508865356
2.3 27.550500869751
};
\addplot [black27]
table {%
2.25 0
2.35 0
};
\addplot [black27]
table {%
2.25 27.550500869751
2.35 27.550500869751
};
\addplot [black27]
table {%
-0.4 0
-0.2 0
};
\addplot [black27]
table {%
0.6 0
0.8 0
};
\addplot [black27]
table {%
1.6 0
1.8 0
};
\addplot [black27]
table {%
-0.2 23.5815000534058
2.77555756156289e-17 23.5815000534058
};
\addplot [black27]
table {%
0.8 17.5064506530762
1 17.5064506530762
};
\addplot [black27]
table {%
1.8 8.16139984130859
2 8.16139984130859
};
\addplot [black27]
table {%
-2.77555756156289e-17 23.1364507675171
0.2 23.1364507675171
};
\addplot [black27]
table {%
1 13.9078006744385
1.2 13.9078006744385
};
\addplot [black27]
table {%
2 8.95014953613281
2.2 8.95014953613281
};
\addplot [black27]
table {%
0.2 13.6730003356934
0.4 13.6730003356934
};
\addplot [black27]
table {%
1.2 10.0893001556396
1.4 10.0893001556396
};
\addplot [black27]
table {%
2.2 6.62629890441895
2.4 6.62629890441895
};
\end{axis}

\end{tikzpicture}
	}
    \subfloat[][DL SINR.]
	{
	    \label{fig:dl_sinr}
\begin{tikzpicture}

\definecolor{black27}{RGB}{27,27,27}
\definecolor{darkgray176}{RGB}{176,176,176}
\definecolor{color4}{RGB}{44,34,57}
\definecolor{color3}{RGB}{124,82,119}
\definecolor{color1}{RGB}{233,211,207}
\definecolor{color2}{RGB}{191,143,160}

\begin{axis}[
width = \textwidth/3.1,
height = 4.35cm,
tick pos=left,
tick align=outside,
x grid style={darkgray176},
xlabel={Rain rate ($\rho$) [mm/h]},
xmajorgrids,
xmin=-0.5, xmax=2.5,
xtick style={color=black},
xtick={0,1,2},
xticklabels={0,15,30},
y grid style={darkgray176},
ylabel={DL SINR [dB]},
ymajorgrids,
ymin=-0.1, ymax=62,
ytick style={color=black},
]

\path [draw=black27, fill=color1]
(axis cs:-0.4,40.416600227356)
--(axis cs:-0.2,40.416600227356)
--(axis cs:-0.2,54.6165990829468)
--(axis cs:-0.4,54.6165990829468)
--(axis cs:-0.4,40.416600227356)
--cycle;
\addplot [black27]
table {%
-0.3 40.416600227356
-0.3 32.293399810791
};
\addplot [black27]
table {%
-0.3 54.6165990829468
-0.3 58.7088012695312
};
\addplot [black27]
table {%
-0.35 32.293399810791
-0.25 32.293399810791
};
\addplot [black27]
table {%
-0.35 58.7088012695312
-0.25 58.7088012695312
};
\path [draw=black27, fill=color1]
(axis cs:0.6,31.9496488571167)
--(axis cs:0.8,31.9496488571167)
--(axis cs:0.8,50.9164991378784)
--(axis cs:0.6,50.9164991378784)
--(axis cs:0.6,31.9496488571167)
--cycle;
\addplot [black27]
table {%
0.7 31.9496488571167
0.7 20.1909999847412
};
\addplot [black27]
table {%
0.7 50.9164991378784
0.7 55.2733993530273
};
\addplot [black27]
table {%
0.65 20.1909999847412
0.75 20.1909999847412
};
\addplot [black27]
table {%
0.65 55.2733993530273
0.75 55.2733993530273
};
\path [draw=black27, fill=color1]
(axis cs:1.6,22.0532252788544)
--(axis cs:1.8,22.0532252788544)
--(axis cs:1.8,46.591700553894)
--(axis cs:1.6,46.591700553894)
--(axis cs:1.6,22.0532252788544)
--cycle;
\addplot [black27]
table {%
1.7 22.0532252788544
1.7 6.04536008834839
};
\addplot [black27]
table {%
1.7 46.591700553894
1.7 51.2579002380371
};
\addplot [black27]
table {%
1.65 6.04536008834839
1.75 6.04536008834839
};
\addplot [black27]
table {%
1.65 51.2579002380371
1.75 51.2579002380371
};
\path [draw=black27, fill=color2]
(axis cs:-0.2,22.3455004692078)
--(axis cs:2.77555756156289e-17,22.3455004692078)
--(axis cs:2.77555756156289e-17,54.6165990829468)
--(axis cs:-0.2,54.6165990829468)
--(axis cs:-0.2,22.3455004692078)
--cycle;
\addplot [black27]
table {%
-0.1 22.3455004692078
-0.1 10.8227996826172
};
\addplot [black27]
table {%
-0.1 54.6165990829468
-0.1 58.7088012695312
};
\addplot [black27]
table {%
-0.15 10.8227996826172
-0.05 10.8227996826172
};
\addplot [black27]
table {%
-0.15 58.7088012695312
-0.05 58.7088012695312
};
\path [draw=black27, fill=color2]
(axis cs:0.8,22.2774252891541)
--(axis cs:1,22.2774252891541)
--(axis cs:1,50.9164991378784)
--(axis cs:0.8,50.9164991378784)
--(axis cs:0.8,22.2774252891541)
--cycle;
\addplot [black27]
table {%
0.9 22.2774252891541
0.9 10.9099998474121
};
\addplot [black27]
table {%
0.9 50.9164991378784
0.9 55.2733993530273
};
\addplot [black27]
table {%
0.85 10.9099998474121
0.95 10.9099998474121
};
\addplot [black27]
table {%
0.85 55.2733993530273
0.95 55.2733993530273
};
\path [draw=black27, fill=color2]
(axis cs:1.8,21.5077257156372)
--(axis cs:2,21.5077257156372)
--(axis cs:2,46.591700553894)
--(axis cs:1.8,46.591700553894)
--(axis cs:1.8,21.5077257156372)
--cycle;
\addplot [black27]
table {%
1.9 21.5077257156372
1.9 10.805100440979
};
\addplot [black27]
table {%
1.9 46.591700553894
1.9 51.2579002380371
};
\addplot [black27]
table {%
1.85 10.805100440979
1.95 10.805100440979
};
\addplot [black27]
table {%
1.85 51.2579002380371
1.95 51.2579002380371
};
\path [draw=black27, fill=color3]
(axis cs:-2.77555756156289e-17,26.5143008232117)
--(axis cs:0.2,26.5143008232117)
--(axis cs:0.2,46.6027240753174)
--(axis cs:-2.77555756156289e-17,46.6027240753174)
--(axis cs:-2.77555756156289e-17,26.5143008232117)
--cycle;
\addplot [black27]
table {%
0.1 26.5143008232117
0.1 18.3143997192383
};
\addplot [black27]
table {%
0.1 46.6027240753174
0.1 58.7088012695312
};
\addplot [black27]
table {%
0.05 18.3143997192383
0.15 18.3143997192383
};
\addplot [black27]
table {%
0.05 58.7088012695312
0.15 58.7088012695312
};
\path [draw=black27, fill=color3]
(axis cs:1,26.7258996963501)
--(axis cs:1.2,26.7258996963501)
--(axis cs:1.2,45.8930997848511)
--(axis cs:1,45.8930997848511)
--(axis cs:1,26.7258996963501)
--cycle;
\addplot [black27]
table {%
1.1 26.7258996963501
1.1 18.0657997131348
};
\addplot [black27]
table {%
1.1 45.8930997848511
1.1 55.2733993530273
};
\addplot [black27]
table {%
1.05 18.0657997131348
1.15 18.0657997131348
};
\addplot [black27]
table {%
1.05 55.2733993530273
1.15 55.2733993530273
};
\path [draw=black27, fill=color3]
(axis cs:2,26.5532746315002)
--(axis cs:2.2,26.5532746315002)
--(axis cs:2.2,41.7435760498047)
--(axis cs:2,41.7435760498047)
--(axis cs:2,26.5532746315002)
--cycle;
\addplot [black27]
table {%
2.1 26.5532746315002
2.1 17.7733001708984
};
\addplot [black27]
table {%
2.1 41.7435760498047
2.1 51.2579002380371
};
\addplot [black27]
table {%
2.05 17.7733001708984
2.15 17.7733001708984
};
\addplot [black27]
table {%
2.05 51.2579002380371
2.15 51.2579002380371
};
\path [draw=black27, fill=color4]
(axis cs:0.2,32.6984758377075)
--(axis cs:0.4,32.6984758377075)
--(axis cs:0.4,43.6048488616943)
--(axis cs:0.2,43.6048488616943)
--(axis cs:0.2,32.6984758377075)
--cycle;
\addplot [black27]
table {%
0.3 32.6984758377075
0.3 16.4820995330811
};
\addplot [black27]
table {%
0.3 43.6048488616943
0.3 55.054500579834
};
\addplot [black27]
table {%
0.25 16.4820995330811
0.35 16.4820995330811
};
\addplot [black27]
table {%
0.25 55.054500579834
0.35 55.054500579834
};
\path [draw=black27, fill=color4]
(axis cs:1.2,32.8506011962891)
--(axis cs:1.4,32.8506011962891)
--(axis cs:1.4,41.5631761550903)
--(axis cs:1.2,41.5631761550903)
--(axis cs:1.2,32.8506011962891)
--cycle;
\addplot [black27]
table {%
1.3 32.8506011962891
1.3 20.3479995727539
};
\addplot [black27]
table {%
1.3 41.5631761550903
1.3 51.4101982116699
};
\addplot [black27]
table {%
1.25 20.3479995727539
1.35 20.3479995727539
};
\addplot [black27]
table {%
1.25 51.4101982116699
1.35 51.4101982116699
};
\path [draw=black27, fill=color4]
(axis cs:2.2,29.2998247146606)
--(axis cs:2.4,29.2998247146606)
--(axis cs:2.4,41.2269992828369)
--(axis cs:2.2,41.2269992828369)
--(axis cs:2.2,29.2998247146606)
--cycle;
\addplot [black27]
table {%
2.3 29.2998247146606
2.3 16.5655994415283
};
\addplot [black27]
table {%
2.3 41.2269992828369
2.3 47.1030006408691
};
\addplot [black27]
table {%
2.25 16.5655994415283
2.35 16.5655994415283
};
\addplot [black27]
table {%
2.25 47.1030006408691
2.35 47.1030006408691
};
\addplot [black27]
table {%
-0.4 45.7103500366211
-0.2 45.7103500366211
};
\addplot [black27]
table {%
0.6 38.8129997253418
0.8 38.8129997253418
};
\addplot [black27]
table {%
1.6 30.7453002929688
1.8 30.7453002929688
};
\addplot [black27]
table {%
-0.2 31.3062000274658
2.77555756156289e-17 31.3062000274658
};
\addplot [black27]
table {%
0.8 31.4521999359131
1 31.4521999359131
};
\addplot [black27]
table {%
1.8 31.4361009597778
2 31.4361009597778
};
\addplot [black27]
table {%
-2.77555756156289e-17 31.7764501571655
0.2 31.7764501571655
};
\addplot [black27]
table {%
1 33.3352012634277
1.2 33.3352012634277
};
\addplot [black27]
table {%
2 34.627799987793
2.2 34.627799987793
};
\addplot [black27]
table {%
0.2 40.6967506408691
0.4 40.6967506408691
};
\addplot [black27]
table {%
1.2 37.6520500183105
1.4 37.6520500183105
};
\addplot [black27]
table {%
2.2 32.9132499694824
2.4 32.9132499694824
};
\end{axis}

\end{tikzpicture}
	}\vskip 0.2cm
    \subfloat[][$\rho= 0$~mm/h.]
	{
	    \label{fig:pdr_r0}
            \input{Figs/PDR_r0}
	}
   \subfloat[][$\rho= 15$~mm/h.]
	{
            \label{fig:pdr_r15}
            \input{Figs/PDR_r15}
	}
    \subfloat[][$\rho= 30$~mm/h.]
	{
            \label{fig:pdr_r30}
            \input{Figs/PDR_r30}
	}\vskip 0.2cm
    \subfloat[][DL latency, $\rho=0$~mm/h.]
	{
	        \label{fig:dl_lat_r0}
            \input{Figs/latency_dl_0_log}
	}
    \subfloat[][UL latency, $\rho=0$~mm/h.]
	{
            \label{fig:ul_lat_r0}
            \input{Figs/latency_ul_0_log}
	}
    \vskip 0.2cm
    \caption{DL SNR, DL interference, DL SINR, PDR, and DL/UL latency, as functions of the source rate for different IAB topologies.}
   \label{fig:rain}
   \vskip -0.1em
\end{figure*}

To test and showcase the proposed 5G NR \gls{iab} simulator, we run a full-stack system-level simulation campaign of a maritime \gls{iab} deployment for remote connectivity. 
In particular, we consider a scenario where coastal base stations, with a fiber connection to the \gls{cn} and the \gls{dn}, play the role of \gls{iab}-donors, and provide a wired endpoint to $8$ \gls{iab}-nodes deployed on moving vessels.
The \gls{iab}-nodes, located near the coastline and moving parallel to it at a velocity of 5~m/s\footnote{We assume a vessel speed of 5~m/s, which is compatible with coastal navigation.
Nevertheless, in \gls{los} conditions, higher speeds are not expected to significantly affect \gls{sinr}, latency, or the overall performance trends.}, operate as UEs and transmit (receive) non-\gls{pdu} \gls{udp} data packets originating from (destined to) their LAN interfaces.

To evaluate the impact of the \gls{iab} network topology on the end-to-end performance, we consider the $4$ different deployment options depicted in~\cref{fig:scenarios}, which are defined based on the maximum depths of \gls{iab}-nodes and the number of \gls{iab}-donors.
In Topology 1, all \gls{iab}-nodes are directly connected to a single \gls{iab}-donor. In Topologies 2 and 3, the maximum \gls{iab}-node depth is 2 and 3, respectively. Topology 4 features two \gls{iab}-donors, and the \gls{iab}-nodes are inter-connected to obtain a maximum depth of 3.

The \gls{iab} network implements in-band wireless backhauling, where the access
and backhaul resources are allocated using either a static \gls{tdm} (unless specified otherwise), or an \gls{fdm} scheme. 
Both \gls{iab}-nodes and \gls{iab}-donors are equipped with a 64-element \gls{upa} antenna, realizing codebook-based analog beamforming.
Each element has a maximum gain of 13~dBi, and the radiation pattern is modeled as in the 3GPP specifications~\cite{3gpp.38.901}.
The \glspl{upa} of the \gls{iab}-donors and all the \glspl{du} of the \gls{iab}-nodes are oriented such that their boresight direction is parallel to the positive X-axis.
In contrast, for the \glspl{mt} of the \gls{iab}-nodes, the orientation is toward their respective parent \gls{iab}-node.\footnote{The impact of the antenna orientation is negligible as long as the target node remains reachable through dynamic beamforming.}
Both IAB-nodes and \gls{iab}-donors operate with a transmission power of 30~dBm, a receiver noise figure of 5~dB, and a bandwidth of 400~MHz.
The full list of simulation parameters is provided in~\cref{tab:parameters}.

In the following, we numerically validate the new features of our IAB simulator described in \cref{sec:simulation-module}, specifically the impact of the rain rate (\cref{sub:rain-rate-eval}), the slot pattern (\cref{sub:slot-eval}), and the MT and DU multiplexing (\cref{sub:multiplexing-eval}).


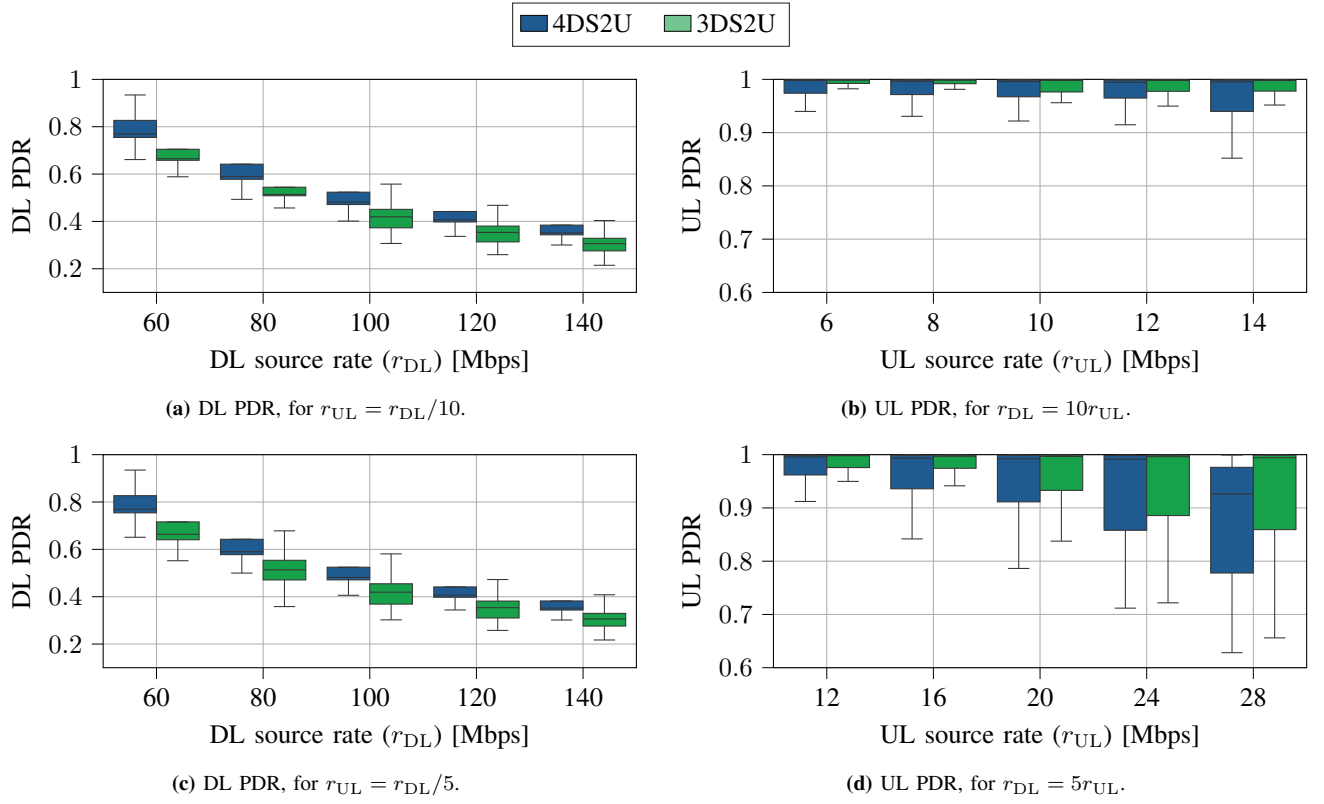
\begin{figure*}[t!]
    \begin{subfigure}[b]{\linewidth}
	\centering
	\setlength\fwidth{\columnwidth}
%
%

\definecolor{color1}{RGB}{31,91,145}
\definecolor{color2}{RGB}{23,161,74}

\begin{tikzpicture}
\pgfplotsset{every tick label/.append style={font=\scriptsize}}

\pgfplotsset{compat=1.11,
	/pgfplots/ybar legend/.style={
		/pgfplots/legend image code/.code={%
			\draw[##1,/tikz/.cd,yshift=-0.25em]
			(0cm,0cm) rectangle (10pt,0.6em);},
	},
}

\begin{axis}[%
width=0,
height=0,
at={(0,0)},
scale only axis,
xtick={},
ytick={},
axis background/.style={fill=white},
legend style={legend cell align=left,
              align=center,
              draw=white!15!black,
              at={(0.5, 1.3)},
              anchor=center,
              /tikz/every even column/.append style={column sep=1em}},
legend columns=2,
]
\addplot[ybar,ybar legend,draw=black,fill=color1,line width=0.08pt]
table[row sep=crcr]{%
	0	0\\
};
\addlegendentry{4DS2U}

\addplot[ybar legend,ybar,draw=black,fill=color2,fill opacity=0.6,line width=0.08pt]
  table[row sep=crcr]{%
	0	0\\
};
\addlegendentry{3DS2U}

\end{axis}
\end{tikzpicture}%
    \end{subfigure}
    \vskip 0.2cm
    \centering
    \subfloat[][DL PDR, for $r_{\rm UL} = r_{\rm DL}/10$.]
	{
	        \label{fig:pdr_0.1_dl}
\begin{tikzpicture}

\definecolor{darkgray176}{RGB}{176,176,176}
\definecolor{color1}{RGB}{31,91,145}
\definecolor{darkslategray53}{RGB}{53,53,53}
\definecolor{color2}{RGB}{23,161,74}

\begin{axis}[
width = \textwidth/2.1,
height = 4.4cm,
tick pos=left,
tick align=outside,
x grid style={darkgray176},
xlabel={DL source rate ($r_{\rm DL}$) [Mbps]},
xmajorgrids,
xmin=-0.5, xmax=4.5,
xtick style={color=black},
xtick={0,1,2,3,4},
xticklabels={60, 80, 100, 120, 140},
y grid style={darkgray176},
ylabel={DL PDR},
ymajorgrids,
ymin=0.1, ymax=1,
ytick style={color=black},
]

\path [draw=darkslategray53, fill=color1]
(axis cs:-0.4,0.754664473684211)
--(axis cs:0,0.754664473684211)
--(axis cs:0,0.826796052631579)
--(axis cs:-0.4,0.826796052631579)
--(axis cs:-0.4,0.754664473684211)
--cycle;
\addplot [darkslategray53]
table {%
-0.2 0.754664473684211
-0.2 0.661263157894737
};
\addplot [darkslategray53]
table {%
-0.2 0.826796052631579
-0.2 0.934184210526316
};
\addplot [darkslategray53]
table {%
-0.3 0.661263157894737
-0.1 0.661263157894737
};
\addplot [darkslategray53]
table {%
-0.3 0.934184210526316
-0.1 0.934184210526316
};
\path [draw=darkslategray53, fill=color1]
(axis cs:0.6,0.577953947368421)
--(axis cs:1,0.577953947368421)
--(axis cs:1,0.641802631578947)
--(axis cs:0.6,0.641802631578947)
--(axis cs:0.6,0.577953947368421)
--cycle;
\addplot [darkslategray53]
table {%
0.8 0.577953947368421
0.8 0.493157894736842
};
\addplot [darkslategray53]
table {%
0.8 0.641802631578947
0.8 0.641802631578947
};
\addplot [darkslategray53]
table {%
0.7 0.493157894736842
0.9 0.493157894736842
};
\addplot [darkslategray53]
table {%
0.7 0.641802631578947
0.9 0.641802631578947
};
\path [draw=darkslategray53, fill=color1]
(axis cs:1.6,0.471263157894737)
--(axis cs:2,0.471263157894737)
--(axis cs:2,0.523467105263158)
--(axis cs:1.6,0.523467105263158)
--(axis cs:1.6,0.471263157894737)
--cycle;
\addplot [darkslategray53]
table {%
1.8 0.471263157894737
1.8 0.401368421052632
};
\addplot [darkslategray53]
table {%
1.8 0.523467105263158
1.8 0.523467105263158
};
\addplot [darkslategray53]
table {%
1.7 0.401368421052632
1.9 0.401368421052632
};
\addplot [darkslategray53]
table {%
1.7 0.523467105263158
1.9 0.523467105263158
};
\path [draw=darkslategray53, fill=color1]
(axis cs:2.6,0.397618421052632)
--(axis cs:3,0.397618421052632)
--(axis cs:3,0.441842105263158)
--(axis cs:2.6,0.441842105263158)
--(axis cs:2.6,0.397618421052632)
--cycle;
\addplot [darkslategray53]
table {%
2.8 0.397618421052632
2.8 0.336710526315789
};
\addplot [darkslategray53]
table {%
2.8 0.441842105263158
2.8 0.441842105263158
};
\addplot [darkslategray53]
table {%
2.7 0.336710526315789
2.9 0.336710526315789
};
\addplot [darkslategray53]
table {%
2.7 0.441842105263158
2.9 0.441842105263158
};
\path [draw=darkslategray53, fill=color1]
(axis cs:3.6,0.343381578947368)
--(axis cs:4,0.343381578947368)
--(axis cs:4,0.384342105263158)
--(axis cs:3.6,0.384342105263158)
--(axis cs:3.6,0.343381578947368)
--cycle;
\addplot [darkslategray53]
table {%
3.8 0.343381578947368
3.8 0.300368421052632
};
\addplot [darkslategray53]
table {%
3.8 0.384342105263158
3.8 0.384342105263158
};
\addplot [darkslategray53]
table {%
3.7 0.300368421052632
3.9 0.300368421052632
};
\addplot [darkslategray53]
table {%
3.7 0.384342105263158
3.9 0.384342105263158
};
\path [draw=darkslategray53, fill=color2]
(axis cs:5.55111512312578e-17,0.65816447368421)
--(axis cs:0.4,0.65816447368421)
--(axis cs:0.4,0.704796052631579)
--(axis cs:5.55111512312578e-17,0.704796052631579)
--(axis cs:5.55111512312578e-17,0.65816447368421)
--cycle;
\addplot [darkslategray53]
table {%
0.2 0.65816447368421
0.2 0.588526315789474
};
\addplot [darkslategray53]
table {%
0.2 0.704796052631579
0.2 0.704796052631579
};
\addplot [darkslategray53]
table {%
0.1 0.588526315789474
0.3 0.588526315789474
};
\addplot [darkslategray53]
table {%
0.1 0.704796052631579
0.3 0.704796052631579
};
\path [draw=darkslategray53, fill=color2]
(axis cs:1,0.50846052631579)
--(axis cs:1.4,0.50846052631579)
--(axis cs:1.4,0.544355263157895)
--(axis cs:1,0.544355263157895)
--(axis cs:1,0.50846052631579)
--cycle;
\addplot [darkslategray53]
table {%
1.2 0.50846052631579
1.2 0.456763157894737
};
\addplot [darkslategray53]
table {%
1.2 0.544355263157895
1.2 0.544355263157895
};
\addplot [darkslategray53]
table {%
1.1 0.456763157894737
1.3 0.456763157894737
};
\addplot [darkslategray53]
table {%
1.1 0.544355263157895
1.3 0.544355263157895
};
\path [draw=darkslategray53, fill=color2]
(axis cs:2,0.372828947368421)
--(axis cs:2.4,0.372828947368421)
--(axis cs:2.4,0.450993421052632)
--(axis cs:2,0.450993421052632)
--(axis cs:2,0.372828947368421)
--cycle;
\addplot [darkslategray53]
table {%
2.2 0.372828947368421
2.2 0.306815789473684
};
\addplot [darkslategray53]
table {%
2.2 0.450993421052632
2.2 0.557578947368421
};
\addplot [darkslategray53]
table {%
2.1 0.306815789473684
2.3 0.306815789473684
};
\addplot [darkslategray53]
table {%
2.1 0.557578947368421
2.3 0.557578947368421
};
\path [draw=darkslategray53, fill=color2]
(axis cs:3,0.313401315789474)
--(axis cs:3.4,0.313401315789474)
--(axis cs:3.4,0.380302631578947)
--(axis cs:3,0.380302631578947)
--(axis cs:3,0.313401315789474)
--cycle;
\addplot [darkslategray53]
table {%
3.2 0.313401315789474
3.2 0.259473684210526
};
\addplot [darkslategray53]
table {%
3.2 0.380302631578947
3.2 0.467921052631579
};
\addplot [darkslategray53]
table {%
3.1 0.259473684210526
3.3 0.259473684210526
};
\addplot [darkslategray53]
table {%
3.1 0.467921052631579
3.3 0.467921052631579
};
\path [draw=darkslategray53, fill=color2]
(axis cs:4,0.275460526315789)
--(axis cs:4.4,0.275460526315789)
--(axis cs:4.4,0.328730263157895)
--(axis cs:4,0.328730263157895)
--(axis cs:4,0.275460526315789)
--cycle;
\addplot [darkslategray53]
table {%
4.2 0.275460526315789
4.2 0.214605263157895
};
\addplot [darkslategray53]
table {%
4.2 0.328730263157895
4.2 0.403578947368421
};
\addplot [darkslategray53]
table {%
4.1 0.214605263157895
4.3 0.214605263157895
};
\addplot [darkslategray53]
table {%
4.1 0.403578947368421
4.3 0.403578947368421
};
\addplot [darkslategray53]
table {%
-0.4 0.769473684210526
0 0.769473684210526
};
\addplot [darkslategray53]
table {%
0.6 0.590131578947368
1 0.590131578947368
};
\addplot [darkslategray53]
table {%
1.6 0.481447368421053
2 0.481447368421053
};
\addplot [darkslategray53]
table {%
2.6 0.406565789473684
3 0.406565789473684
};
\addplot [darkslategray53]
table {%
3.6 0.351815789473684
4 0.351815789473684
};
\addplot [darkslategray53]
table {%
5.55111512312578e-17 0.664881578947368
0.4 0.664881578947368
};
\addplot [darkslategray53]
table {%
1 0.514131578947368
1.4 0.514131578947368
};
\addplot [darkslategray53]
table {%
2 0.419131578947368
2.4 0.419131578947368
};
\addplot [darkslategray53]
table {%
3 0.353592105263158
3.4 0.353592105263158
};
\addplot [darkslategray53]
table {%
4 0.305855263157895
4.4 0.305855263157895
};
\end{axis}

\end{tikzpicture}
	}
    \subfloat[][UL PDR, for $r_{\rm DL} = 10r_{\rm UL}$.]
	{
            \label{fig:pdr_0.1_ul}
\begin{tikzpicture}

\definecolor{darkgray176}{RGB}{176,176,176}
\definecolor{color1}{RGB}{31,91,145}
\definecolor{darkslategray53}{RGB}{53,53,53}
\definecolor{color2}{RGB}{23,161,74}

\begin{axis}[
width = \textwidth/2.1,
height = 4.4cm,
tick pos=left,
tick align=outside,
x grid style={darkgray176},
xlabel={UL source rate ($r_{\rm UL}$) [Mbps]},
xmajorgrids,
xmin=-0.5, xmax=4.5,
xtick style={color=black},
xtick={0,1,2,3,4},
xticklabels={6,8,10,12,14},
y grid style={darkgray176},
ylabel={UL PDR},
ymajorgrids,
ymin=0.6, ymax=1,
ytick style={color=black},
]

\path [draw=darkslategray53, fill=color1]
(axis cs:-0.4,0.973960526315789)
--(axis cs:0,0.973960526315789)
--(axis cs:0,0.998868421052632)
--(axis cs:-0.4,0.998868421052632)
--(axis cs:-0.4,0.973960526315789)
--cycle;
\addplot [darkslategray53]
table {%
-0.2 0.973960526315789
-0.2 0.939789473684211
};
\addplot [darkslategray53]
table {%
-0.2 0.998868421052632
-0.2 0.999289473684211
};
\addplot [darkslategray53]
table {%
-0.3 0.939789473684211
-0.1 0.939789473684211
};
\addplot [darkslategray53]
table {%
-0.3 0.999289473684211
-0.1 0.999289473684211
};
\path [draw=darkslategray53, fill=color1]
(axis cs:0.6,0.971414473684211)
--(axis cs:1,0.971414473684211)
--(axis cs:1,0.998842105263158)
--(axis cs:0.6,0.998842105263158)
--(axis cs:0.6,0.971414473684211)
--cycle;
\addplot [darkslategray53]
table {%
0.8 0.971414473684211
0.8 0.930631578947368
};
\addplot [darkslategray53]
table {%
0.8 0.998842105263158
0.8 0.999289473684211
};
\addplot [darkslategray53]
table {%
0.7 0.930631578947368
0.9 0.930631578947368
};
\addplot [darkslategray53]
table {%
0.7 0.999289473684211
0.9 0.999289473684211
};
\path [draw=darkslategray53, fill=color1]
(axis cs:1.6,0.967309210526316)
--(axis cs:2,0.967309210526316)
--(axis cs:2,0.998842105263158)
--(axis cs:1.6,0.998842105263158)
--(axis cs:1.6,0.967309210526316)
--cycle;
\addplot [darkslategray53]
table {%
1.8 0.967309210526316
1.8 0.921815789473684
};
\addplot [darkslategray53]
table {%
1.8 0.998842105263158
1.8 0.999289473684211
};
\addplot [darkslategray53]
table {%
1.7 0.921815789473684
1.9 0.921815789473684
};
\addplot [darkslategray53]
table {%
1.7 0.999289473684211
1.9 0.999289473684211
};
\path [draw=darkslategray53, fill=color1]
(axis cs:2.6,0.964855263157895)
--(axis cs:3,0.964855263157895)
--(axis cs:3,0.998842105263158)
--(axis cs:2.6,0.998842105263158)
--(axis cs:2.6,0.964855263157895)
--cycle;
\addplot [darkslategray53]
table {%
2.8 0.964855263157895
2.8 0.914736842105263
};
\addplot [darkslategray53]
table {%
2.8 0.998842105263158
2.8 0.999289473684211
};
\addplot [darkslategray53]
table {%
2.7 0.914736842105263
2.9 0.914736842105263
};
\addplot [darkslategray53]
table {%
2.7 0.999289473684211
2.9 0.999289473684211
};
\path [draw=darkslategray53, fill=color1]
(axis cs:3.6,0.939815789473684)
--(axis cs:4,0.939815789473684)
--(axis cs:4,0.998842105263158)
--(axis cs:3.6,0.998842105263158)
--(axis cs:3.6,0.939815789473684)
--cycle;
\addplot [darkslategray53]
table {%
3.8 0.939815789473684
3.8 0.852052631578947
};
\addplot [darkslategray53]
table {%
3.8 0.998842105263158
3.8 0.999289473684211
};
\addplot [darkslategray53]
table {%
3.7 0.852052631578947
3.9 0.852052631578947
};
\addplot [darkslategray53]
table {%
3.7 0.999289473684211
3.9 0.999289473684211
};
\path [draw=darkslategray53, fill=color2]
(axis cs:5.55111512312578e-17,0.992368421052632)
--(axis cs:0.4,0.992368421052632)
--(axis cs:0.4,0.999236842105263)
--(axis cs:5.55111512312578e-17,0.999236842105263)
--(axis cs:5.55111512312578e-17,0.992368421052632)
--cycle;
\addplot [darkslategray53]
table {%
0.2 0.992368421052632
0.2 0.98228947368421
};
\addplot [darkslategray53]
table {%
0.2 0.999236842105263
0.2 0.999289473684211
};
\addplot [darkslategray53]
table {%
0.1 0.98228947368421
0.3 0.98228947368421
};
\addplot [darkslategray53]
table {%
0.1 0.999289473684211
0.3 0.999289473684211
};
\path [draw=darkslategray53, fill=color2]
(axis cs:1,0.991855263157895)
--(axis cs:1.4,0.991855263157895)
--(axis cs:1.4,0.999236842105263)
--(axis cs:1,0.999236842105263)
--(axis cs:1,0.991855263157895)
--cycle;
\addplot [darkslategray53]
table {%
1.2 0.991855263157895
1.2 0.98128947368421
};
\addplot [darkslategray53]
table {%
1.2 0.999236842105263
1.2 0.999289473684211
};
\addplot [darkslategray53]
table {%
1.1 0.98128947368421
1.3 0.98128947368421
};
\addplot [darkslategray53]
table {%
1.1 0.999289473684211
1.3 0.999289473684211
};
\path [draw=darkslategray53, fill=color2]
(axis cs:2,0.976328947368421)
--(axis cs:2.4,0.976328947368421)
--(axis cs:2.4,0.998973684210526)
--(axis cs:2,0.998973684210526)
--(axis cs:2,0.976328947368421)
--cycle;
\addplot [darkslategray53]
table {%
2.2 0.976328947368421
2.2 0.956157894736842
};
\addplot [darkslategray53]
table {%
2.2 0.998973684210526
2.2 0.999289473684211
};
\addplot [darkslategray53]
table {%
2.1 0.956157894736842
2.3 0.956157894736842
};
\addplot [darkslategray53]
table {%
2.1 0.999289473684211
2.3 0.999289473684211
};
\path [draw=darkslategray53, fill=color2]
(axis cs:3,0.97758552631579)
--(axis cs:3.4,0.97758552631579)
--(axis cs:3.4,0.998973684210526)
--(axis cs:3,0.998973684210526)
--(axis cs:3,0.97758552631579)
--cycle;
\addplot [darkslategray53]
table {%
3.2 0.97758552631579
3.2 0.949789473684211
};
\addplot [darkslategray53]
table {%
3.2 0.998973684210526
3.2 0.999289473684211
};
\addplot [darkslategray53]
table {%
3.1 0.949789473684211
3.3 0.949789473684211
};
\addplot [darkslategray53]
table {%
3.1 0.999289473684211
3.3 0.999289473684211
};
\path [draw=darkslategray53, fill=color2]
(axis cs:4,0.977914473684211)
--(axis cs:4.4,0.977914473684211)
--(axis cs:4.4,0.998973684210526)
--(axis cs:4,0.998973684210526)
--(axis cs:4,0.977914473684211)
--cycle;
\addplot [darkslategray53]
table {%
4.2 0.977914473684211
4.2 0.951789473684211
};
\addplot [darkslategray53]
table {%
4.2 0.998973684210526
4.2 0.999289473684211
};
\addplot [darkslategray53]
table {%
4.1 0.951789473684211
4.3 0.951789473684211
};
\addplot [darkslategray53]
table {%
4.1 0.999289473684211
4.3 0.999289473684211
};
\addplot [darkslategray53]
table {%
-0.4 0.997986842105263
0 0.997986842105263
};
\addplot [darkslategray53]
table {%
0.6 0.996276315789474
1 0.996276315789474
};
\addplot [darkslategray53]
table {%
1.6 0.995815789473684
2 0.995815789473684
};
\addplot [darkslategray53]
table {%
2.6 0.994473684210526
3 0.994473684210526
};
\addplot [darkslategray53]
table {%
3.6 0.995355263157895
4 0.995355263157895
};
\addplot [darkslategray53]
table {%
5.55111512312578e-17 0.998578947368421
0.4 0.998578947368421
};
\addplot [darkslategray53]
table {%
1 0.998578947368421
1.4 0.998578947368421
};
\addplot [darkslategray53]
table {%
2 0.998210526315789
2.4 0.998210526315789
};
\addplot [darkslategray53]
table {%
3 0.997960526315789
3.4 0.997960526315789
};
\addplot [darkslategray53]
table {%
4 0.997842105263158
4.4 0.997842105263158
};
\end{axis}

\end{tikzpicture}
	}
    \vskip 0.2cm
    \subfloat[][DL PDR, for $r_{\rm UL} = r_{\rm DL}/5$.]
	{
            \label{fig:pdr_0.2_dl}
\begin{tikzpicture}

\definecolor{darkgray176}{RGB}{176,176,176}
\definecolor{color1}{RGB}{31,91,145}
\definecolor{darkslategray53}{RGB}{53,53,53}
\definecolor{color2}{RGB}{23,161,74}

\begin{axis}[
width = \textwidth/2.1,
height = 4.4cm,
tick pos=left,
tick align=outside,
x grid style={darkgray176},
xlabel={DL source rate ($r_{\rm DL}$) [Mbps]},
xmajorgrids,
xmin=-0.5, xmax=4.5,
xtick style={color=black},
xtick={0,1,2,3,4},
xticklabels={60, 80, 100, 120, 140},
y grid style={darkgray176},
ylabel={DL PDR},
ymajorgrids,
ymin=0.1, ymax=1,
ytick style={color=black}
]

\path [draw=darkslategray53, fill=color1]
(axis cs:-0.4,0.75425)
--(axis cs:0,0.75425)
--(axis cs:0,0.826763157894737)
--(axis cs:-0.4,0.826763157894737)
--(axis cs:-0.4,0.75425)
--cycle;
\addplot [darkslategray53]
table {%
-0.2 0.75425
-0.2 0.651078947368421
};
\addplot [darkslategray53]
table {%
-0.2 0.826763157894737
-0.2 0.934868421052632
};
\addplot [darkslategray53]
table {%
-0.3 0.651078947368421
-0.1 0.651078947368421
};
\addplot [darkslategray53]
table {%
-0.3 0.934868421052632
-0.1 0.934868421052632
};
\path [draw=darkslategray53, fill=color1]
(axis cs:0.6,0.577796052631579)
--(axis cs:1,0.577796052631579)
--(axis cs:1,0.642651315789474)
--(axis cs:0.6,0.642651315789474)
--(axis cs:0.6,0.577796052631579)
--cycle;
\addplot [darkslategray53]
table {%
0.8 0.577796052631579
0.8 0.499763157894737
};
\addplot [darkslategray53]
table {%
0.8 0.642651315789474
0.8 0.642651315789474
};
\addplot [darkslategray53]
table {%
0.7 0.499763157894737
0.9 0.499763157894737
};
\addplot [darkslategray53]
table {%
0.7 0.642651315789474
0.9 0.642651315789474
};
\path [draw=darkslategray53, fill=color1]
(axis cs:1.6,0.471065789473684)
--(axis cs:2,0.471065789473684)
--(axis cs:2,0.524559210526316)
--(axis cs:1.6,0.524559210526316)
--(axis cs:1.6,0.471065789473684)
--cycle;
\addplot [darkslategray53]
table {%
1.8 0.471065789473684
1.8 0.406
};
\addplot [darkslategray53]
table {%
1.8 0.524559210526316
1.8 0.524559210526316
};
\addplot [darkslategray53]
table {%
1.7 0.406
1.9 0.406
};
\addplot [darkslategray53]
table {%
1.7 0.524559210526316
1.9 0.524559210526316
};
\path [draw=darkslategray53, fill=color1]
(axis cs:2.6,0.397618421052632)
--(axis cs:3,0.397618421052632)
--(axis cs:3,0.441309210526316)
--(axis cs:2.6,0.441309210526316)
--(axis cs:2.6,0.397618421052632)
--cycle;
\addplot [darkslategray53]
table {%
2.8 0.397618421052632
2.8 0.344
};
\addplot [darkslategray53]
table {%
2.8 0.441309210526316
2.8 0.441309210526316
};
\addplot [darkslategray53]
table {%
2.7 0.344
2.9 0.344
};
\addplot [darkslategray53]
table {%
2.7 0.441309210526316
2.9 0.441309210526316
};
\path [draw=darkslategray53, fill=color1]
(axis cs:3.6,0.343625)
--(axis cs:4,0.343625)
--(axis cs:4,0.382019736842105)
--(axis cs:3.6,0.382019736842105)
--(axis cs:3.6,0.343625)
--cycle;
\addplot [darkslategray53]
table {%
3.8 0.343625
3.8 0.301526315789474
};
\addplot [darkslategray53]
table {%
3.8 0.382019736842105
3.8 0.382019736842105
};
\addplot [darkslategray53]
table {%
3.7 0.301526315789474
3.9 0.301526315789474
};
\addplot [darkslategray53]
table {%
3.7 0.382019736842105
3.9 0.382019736842105
};
\path [draw=darkslategray53, fill=color2]
(axis cs:5.55111512312578e-17,0.640605263157895)
--(axis cs:0.4,0.640605263157895)
--(axis cs:0.4,0.716065789473684)
--(axis cs:5.55111512312578e-17,0.716065789473684)
--(axis cs:5.55111512312578e-17,0.640605263157895)
--cycle;
\addplot [darkslategray53]
table {%
0.2 0.640605263157895
0.2 0.551947368421053
};
\addplot [darkslategray53]
table {%
0.2 0.716065789473684
0.2 0.716065789473684
};
\addplot [darkslategray53]
table {%
0.1 0.551947368421053
0.3 0.551947368421053
};
\addplot [darkslategray53]
table {%
0.1 0.716065789473684
0.3 0.716065789473684
};
\path [draw=darkslategray53, fill=color2]
(axis cs:1,0.470809210526316)
--(axis cs:1.4,0.470809210526316)
--(axis cs:1.4,0.553848684210526)
--(axis cs:1,0.553848684210526)
--(axis cs:1,0.470809210526316)
--cycle;
\addplot [darkslategray53]
table {%
1.2 0.470809210526316
1.2 0.358263157894737
};
\addplot [darkslategray53]
table {%
1.2 0.553848684210526
1.2 0.678184210526316
};
\addplot [darkslategray53]
table {%
1.1 0.358263157894737
1.3 0.358263157894737
};
\addplot [darkslategray53]
table {%
1.1 0.678184210526316
1.3 0.678184210526316
};
\path [draw=darkslategray53, fill=color2]
(axis cs:2,0.368710526315789)
--(axis cs:2.4,0.368710526315789)
--(axis cs:2.4,0.454631578947368)
--(axis cs:2,0.454631578947368)
--(axis cs:2,0.368710526315789)
--cycle;
\addplot [darkslategray53]
table {%
2.2 0.368710526315789
2.2 0.302184210526316
};
\addplot [darkslategray53]
table {%
2.2 0.454631578947368
2.2 0.580894736842105
};
\addplot [darkslategray53]
table {%
2.1 0.302184210526316
2.3 0.302184210526316
};
\addplot [darkslategray53]
table {%
2.1 0.580894736842105
2.3 0.580894736842105
};
\path [draw=darkslategray53, fill=color2]
(axis cs:3,0.310059210526316)
--(axis cs:3.4,0.310059210526316)
--(axis cs:3.4,0.381519736842105)
--(axis cs:3,0.381519736842105)
--(axis cs:3,0.310059210526316)
--cycle;
\addplot [darkslategray53]
table {%
3.2 0.310059210526316
3.2 0.257605263157895
};
\addplot [darkslategray53]
table {%
3.2 0.381519736842105
3.2 0.472473684210526
};
\addplot [darkslategray53]
table {%
3.1 0.257605263157895
3.3 0.257605263157895
};
\addplot [darkslategray53]
table {%
3.1 0.472473684210526
3.3 0.472473684210526
};
\path [draw=darkslategray53, fill=color2]
(axis cs:4,0.276046052631579)
--(axis cs:4.4,0.276046052631579)
--(axis cs:4.4,0.329532894736842)
--(axis cs:4,0.329532894736842)
--(axis cs:4,0.276046052631579)
--cycle;
\addplot [darkslategray53]
table {%
4.2 0.276046052631579
4.2 0.217131578947368
};
\addplot [darkslategray53]
table {%
4.2 0.329532894736842
4.2 0.408105263157895
};
\addplot [darkslategray53]
table {%
4.1 0.217131578947368
4.3 0.217131578947368
};
\addplot [darkslategray53]
table {%
4.1 0.408105263157895
4.3 0.408105263157895
};
\addplot [darkslategray53]
table {%
-0.4 0.769355263157895
0 0.769355263157895
};
\addplot [darkslategray53]
table {%
0.6 0.590223684210526
1 0.590223684210526
};
\addplot [darkslategray53]
table {%
1.6 0.481486842105263
2 0.481486842105263
};
\addplot [darkslategray53]
table {%
2.6 0.406552631578947
3 0.406552631578947
};
\addplot [darkslategray53]
table {%
3.6 0.351842105263158
4 0.351842105263158
};
\addplot [darkslategray53]
table {%
5.55111512312578e-17 0.663776315789474
0.4 0.663776315789474
};
\addplot [darkslategray53]
table {%
1 0.513473684210526
1.4 0.513473684210526
};
\addplot [darkslategray53]
table {%
2 0.419157894736842
2.4 0.419157894736842
};
\addplot [darkslategray53]
table {%
3 0.353592105263158
3.4 0.353592105263158
};
\addplot [darkslategray53]
table {%
4 0.305894736842105
4.4 0.305894736842105
};
\end{axis}

\end{tikzpicture}
	}
    \subfloat[][UL PDR, for $r_{\rm DL} = 5r_{\rm UL}$.]
	{
            \label{fig:pdr_0.2_ul}
\begin{tikzpicture}

\definecolor{darkgray176}{RGB}{176,176,176}
\definecolor{color1}{RGB}{31,91,145}
\definecolor{darkslategray53}{RGB}{53,53,53}
\definecolor{color2}{RGB}{23,161,74}

\begin{axis}[
width = \textwidth/2.1,
height = 4.4cm,
tick pos=left,
tick align=outside,
x grid style={darkgray176},
xlabel={UL source rate ($r_{\rm UL}$) [Mbps]},
xmajorgrids,
xmin=-0.5, xmax=4.5,
xtick style={color=black},
xtick={0,1,2,3,4},
xticklabels={12, 16, 20, 24, 28},
y grid style={darkgray176},
ylabel={UL PDR},
ymajorgrids,
ymin=0.6, ymax=1,
ytick style={color=black},
]

\path [draw=darkslategray53, fill=color1]
(axis cs:-0.4,0.961875)
--(axis cs:0,0.961875)
--(axis cs:0,0.998842105263158)
--(axis cs:-0.4,0.998842105263158)
--(axis cs:-0.4,0.961875)
--cycle;
\addplot [darkslategray53]
table {%
-0.2 0.961875
-0.2 0.912105263157895
};
\addplot [darkslategray53]
table {%
-0.2 0.998842105263158
-0.2 0.999289473684211
};
\addplot [darkslategray53]
table {%
-0.3 0.912105263157895
-0.1 0.912105263157895
};
\addplot [darkslategray53]
table {%
-0.3 0.999289473684211
-0.1 0.999289473684211
};
\path [draw=darkslategray53, fill=color1]
(axis cs:0.6,0.935901315789474)
--(axis cs:1,0.935901315789474)
--(axis cs:1,0.998842105263158)
--(axis cs:0.6,0.998842105263158)
--(axis cs:0.6,0.935901315789474)
--cycle;
\addplot [darkslategray53]
table {%
0.8 0.935901315789474
0.8 0.841815789473684
};
\addplot [darkslategray53]
table {%
0.8 0.998842105263158
0.8 0.999289473684211
};
\addplot [darkslategray53]
table {%
0.7 0.841815789473684
0.9 0.841815789473684
};
\addplot [darkslategray53]
table {%
0.7 0.999289473684211
0.9 0.999289473684211
};
\path [draw=darkslategray53, fill=color1]
(axis cs:1.6,0.911335526315789)
--(axis cs:2,0.911335526315789)
--(axis cs:2,0.998842105263158)
--(axis cs:1.6,0.998842105263158)
--(axis cs:1.6,0.911335526315789)
--cycle;
\addplot [darkslategray53]
table {%
1.8 0.911335526315789
1.8 0.786368421052632
};
\addplot [darkslategray53]
table {%
1.8 0.998842105263158
1.8 0.999289473684211
};
\addplot [darkslategray53]
table {%
1.7 0.786368421052632
1.9 0.786368421052632
};
\addplot [darkslategray53]
table {%
1.7 0.999289473684211
1.9 0.999289473684211
};
\path [draw=darkslategray53, fill=color1]
(axis cs:2.6,0.858006578947368)
--(axis cs:3,0.858006578947368)
--(axis cs:3,0.998368421052632)
--(axis cs:2.6,0.998368421052632)
--(axis cs:2.6,0.858006578947368)
--cycle;
\addplot [darkslategray53]
table {%
2.8 0.858006578947368
2.8 0.711842105263158
};
\addplot [darkslategray53]
table {%
2.8 0.998368421052632
2.8 0.999289473684211
};
\addplot [darkslategray53]
table {%
2.7 0.711842105263158
2.9 0.711842105263158
};
\addplot [darkslategray53]
table {%
2.7 0.999289473684211
2.9 0.999289473684211
};
\path [draw=darkslategray53, fill=color1]
(axis cs:3.6,0.777802631578947)
--(axis cs:4,0.777802631578947)
--(axis cs:4,0.976256578947368)
--(axis cs:3.6,0.976256578947368)
--(axis cs:3.6,0.777802631578947)
--cycle;
\addplot [darkslategray53]
table {%
3.8 0.777802631578947
3.8 0.628184210526316
};
\addplot [darkslategray53]
table {%
3.8 0.976256578947368
3.8 0.999289473684211
};
\addplot [darkslategray53]
table {%
3.7 0.628184210526316
3.9 0.628184210526316
};
\addplot [darkslategray53]
table {%
3.7 0.999289473684211
3.9 0.999289473684211
};
\path [draw=darkslategray53, fill=color2]
(axis cs:5.55111512312578e-17,0.975723684210526)
--(axis cs:0.4,0.975723684210526)
--(axis cs:0.4,0.998973684210526)
--(axis cs:5.55111512312578e-17,0.998973684210526)
--(axis cs:5.55111512312578e-17,0.975723684210526)
--cycle;
\addplot [darkslategray53]
table {%
0.2 0.975723684210526
0.2 0.949789473684211
};
\addplot [darkslategray53]
table {%
0.2 0.998973684210526
0.2 0.999289473684211
};
\addplot [darkslategray53]
table {%
0.1 0.949789473684211
0.3 0.949789473684211
};
\addplot [darkslategray53]
table {%
0.1 0.999289473684211
0.3 0.999289473684211
};
\path [draw=darkslategray53, fill=color2]
(axis cs:1,0.974434210526316)
--(axis cs:1.4,0.974434210526316)
--(axis cs:1.4,0.998973684210526)
--(axis cs:1,0.998973684210526)
--(axis cs:1,0.974434210526316)
--cycle;
\addplot [darkslategray53]
table {%
1.2 0.974434210526316
1.2 0.941526315789474
};
\addplot [darkslategray53]
table {%
1.2 0.998973684210526
1.2 0.999289473684211
};
\addplot [darkslategray53]
table {%
1.1 0.941526315789474
1.3 0.941526315789474
};
\addplot [darkslategray53]
table {%
1.1 0.999289473684211
1.3 0.999289473684211
};
\path [draw=darkslategray53, fill=color2]
(axis cs:2,0.932907894736842)
--(axis cs:2.4,0.932907894736842)
--(axis cs:2.4,0.998973684210526)
--(axis cs:2,0.998973684210526)
--(axis cs:2,0.932907894736842)
--cycle;
\addplot [darkslategray53]
table {%
2.2 0.932907894736842
2.2 0.837631578947368
};
\addplot [darkslategray53]
table {%
2.2 0.998973684210526
2.2 0.999289473684211
};
\addplot [darkslategray53]
table {%
2.1 0.837631578947368
2.3 0.837631578947368
};
\addplot [darkslategray53]
table {%
2.1 0.999289473684211
2.3 0.999289473684211
};
\path [draw=darkslategray53, fill=color2]
(axis cs:3,0.885638157894737)
--(axis cs:3.4,0.885638157894737)
--(axis cs:3.4,0.998973684210526)
--(axis cs:3,0.998973684210526)
--(axis cs:3,0.885638157894737)
--cycle;
\addplot [darkslategray53]
table {%
3.2 0.885638157894737
3.2 0.721842105263158
};
\addplot [darkslategray53]
table {%
3.2 0.998973684210526
3.2 0.999289473684211
};
\addplot [darkslategray53]
table {%
3.1 0.721842105263158
3.3 0.721842105263158
};
\addplot [darkslategray53]
table {%
3.1 0.999289473684211
3.3 0.999289473684211
};
\path [draw=darkslategray53, fill=color2]
(axis cs:4,0.859217105263158)
--(axis cs:4.4,0.859217105263158)
--(axis cs:4.4,0.998894736842105)
--(axis cs:4,0.998894736842105)
--(axis cs:4,0.859217105263158)
--cycle;
\addplot [darkslategray53]
table {%
4.2 0.859217105263158
4.2 0.656026315789474
};
\addplot [darkslategray53]
table {%
4.2 0.998894736842105
4.2 0.999289473684211
};
\addplot [darkslategray53]
table {%
4.1 0.656026315789474
4.3 0.656026315789474
};
\addplot [darkslategray53]
table {%
4.1 0.999289473684211
4.3 0.999289473684211
};
\addplot [darkslategray53]
table {%
-0.4 0.995763157894737
0 0.995763157894737
};
\addplot [darkslategray53]
table {%
0.6 0.993263157894737
1 0.993263157894737
};
\addplot [darkslategray53]
table {%
1.6 0.992276315789474
2 0.992276315789474
};
\addplot [darkslategray53]
table {%
2.6 0.991460526315789
3 0.991460526315789
};
\addplot [darkslategray53]
table {%
3.6 0.926171052631579
4 0.926171052631579
};
\addplot [darkslategray53]
table {%
5.55111512312578e-17 0.997960526315789
0.4 0.997960526315789
};
\addplot [darkslategray53]
table {%
1 0.996947368421053
1.4 0.996947368421053
};
\addplot [darkslategray53]
table {%
2 0.996723684210526
2.4 0.996723684210526
};
\addplot [darkslategray53]
table {%
3 0.996092105263158
3.4 0.996092105263158
};
\addplot [darkslategray53]
table {%
4 0.994197368421053
4.4 0.994197368421053
};
\end{axis}

\end{tikzpicture}
	}
    \vskip 0.2cm
    \caption{DL and UL PDR as a function of the source rate for different slot patterns in Topology 3.}
   \label{fig:pdr}
   \vskip -0.1em
\end{figure*}

\subsection{Rain Rate}
\label{sub:rain-rate-eval}
First, we assess how the rain rate affects the IAB network performance for different topologies. 
In~\cref{fig:dl_snr,fig:dl_sinr}, we compare the \gls{snr} and the \gls{sinr}, respectively, under different meteorological conditions, ranging from clear weather to rain rates of up to $\rho=30$~mm/h.

In Topology 1, all \gls{iab}-nodes are directly connected to the \gls{iab}-donor.
This configuration results in relatively long (i.e., up to approximately 3~km) child-to-parent links, which have a negative effect on the \gls{snr}. On the contrary, it guarantees an interference-free deployment since all transmissions are centrally coordinated by the \gls{iab}-donor through a \gls{tdma}-based scheduling scheme. 
In contrast, Topologies 2 and 3 support multi-hop connections, where \gls{iab}-nodes can act either as a parent or as a child, thus reducing the length of the link and improving the average backhaul \gls{snr} compared to Topology 1 (up to $+15$ dB).
However, as described in~\cref{sec:duplexing}, this configuration can create concurrent transmissions among \gls{iab}-nodes of the same layer (even or odd). Therefore, the resulting median interference is approximately 25~dB under clear weather conditions (see \cref{fig:dl_int}, for $\rho=0$ mm/h).
Therefore, despite the longer links, Topology 1 achieves a higher \gls{sinr} than Topologies 2 and 3.
Finally, in Topology 4, the presence of two \gls{iab}-donors effectively creates two separate yet potentially interfering \gls{iab} networks, which may be a source of strong interference in certain conditions, as demonstrated by the high upper whiskers in \cref{fig:dl_int} (up to more than 40 dB for $\rho=0$ mm/h).

As discussed in~\cref{sec:rain}, rain attenuation is directly proportional to the link length.
In \cref{fig:dl_sinr} we see that Topology 1 achieves the highest \gls{sinr} without rain given the absence of interference.
However, at $\rho=15$~mm/h, the minimum \gls{sinr} decreases by more than 10~dB, and at $\rho=30$~mm/h the median \gls{sinr} is even lower than in Topologies 2 and 3 due to the effect of the long links.
Interestingly, for Topologies 2, 3, and 4, rainfall can successfully mitigate interference, as interfering paths are generally longer than the intended communication links, and thus experience greater attenuation. For example, the median SINR in Topology 3 improves by around 5~dB from $\rho=0$~mm/h to $\rho=30$~mm/h.

\cref{fig:pdr_r0,fig:pdr_r15,fig:pdr_r30} show the \gls{pdr} as a function of the \gls{dl} source rate $r_{\rm DL}$ and rain rate $\rho$.
The traffic is mainly \gls{dl}, where each \gls{iab}-node receives data with a \gls{dl} source rate $r_{\rm DL}$, and transmits data with a \gls{ul} source rate $r_{\rm UL} = r_{\rm DL}/5$.
As expected, the \gls{pdr} decreases as the source rate increases due to possible network congestion.
Without rain, Topology 1 outperforms Topologies 2 and 3. Nevertheless, Topology 4 achieves the best performance, with a \gls{pdr} of 1 with up to $r_{\rm DL}=100$~Mbps, given the presence of an additional \gls{iab}-donor and additional available resources.
As previously discussed, rain has a negative impact on Topology 1, 
while for Topologies 2, 3, and 4, rain attenuation actually helps reduce the effect of interference. 

Finally, in~\cref{fig:dl_lat_r0,fig:ul_lat_r0}, we evaluate the \gls{dl} and \gls{ul} latency under clear weather conditions using a 4DS2U slot pattern, meaning that each scheduling period consists of four consecutive \gls{dl} slots, followed by a switch slot, and then two \gls{ul} slots.
The latency is defined as the time from when a packet is generated at the application layer of the local host to when it is received at the remote host (i.e., from the IAB-donor to the IAB-node), and accounts for both transmission and queuing delays.
We use a \gls{rr} scheduler, ensuring fair and balanced distribution of \gls{ofdm} symbols across all \gls{iab}-nodes.
We observe that the latency trends are similar to those of the \gls{pdr} in~\cref{fig:pdr_r0}. 
As the source rate increases, the system approaches its capacity limits, and the end-to-end latency is dominated by the resulting queuing delays. 
Although the 4DS2U pattern allocates fewer symbols to \gls{ul} transmissions, the \gls{ul} source rate is five times lower than the \gls{dl} source rate.
Consequently, \gls{ul} queues are less congested than in \gls{dl}, resulting in lower \gls{ul} latency.
Traffic accumulation occurs primarily at \gls{iab}-nodes closer to the donor, where data traffic from multiple upstream nodes converges as it propagates toward the donor, and therefore experiences higher queuing delay as the source rate increases.
For example, in Topology 3, the median \gls{dl} latency increases from approximately 100~ms at $r_{\rm DL}=60$~Mbps to 500~ms at $r_{\rm DL}=140$~Mbps.
The introduction of a second \gls{iab}-donor in Topology 4 increases the network capacity, reducing both \gls{dl} and \gls{ul} delays.
In this configuration, the \gls{dl} latency remains below 10~ms for source rates up to 100~Mbps.

\subsection{Slot Pattern}
\label{sub:slot-eval}
In this section, we focus on Topology 3, and evaluate the impact of two different slot patterns vs. the source rate.
Specifically, the 4DS2U slot pattern allocates 4 \gls{dl} slots and 2 \gls{ul} slots within each set of 7 consecutive slots, while the 3DS2U slot pattern allocates 3 \gls{dl} slots and 2 \gls{ul} slots within each set of 6 consecutive slots.
In~\cref{fig:pdr_0.1_dl,fig:pdr_0.1_ul}, we set $r_{\rm UL} = r_{\rm DL}/10$, and observe that the 4DS2U configuration improves the \gls{dl} \gls{pdr} compared to the 3DS2U configuration (0.75 vs. 0.65 for $r_{\rm DL}=60$ Mbps), while for the \gls{ul} \gls{pdr} it is the opposite, given that the system configures more \gls{dl} slots and transmission opportunities.
In~\cref{fig:pdr_0.2_dl,fig:pdr_0.2_ul}, we increase the \gls{ul} source rate to $r_{\rm UL} = r_{\rm DL}/5$, and see that the \gls{ul} \gls{pdr} further decreases (by up to around 20\%), given the insufficient number of UL resources to handle the additional UL traffic.

\begin{figure*}[t!]
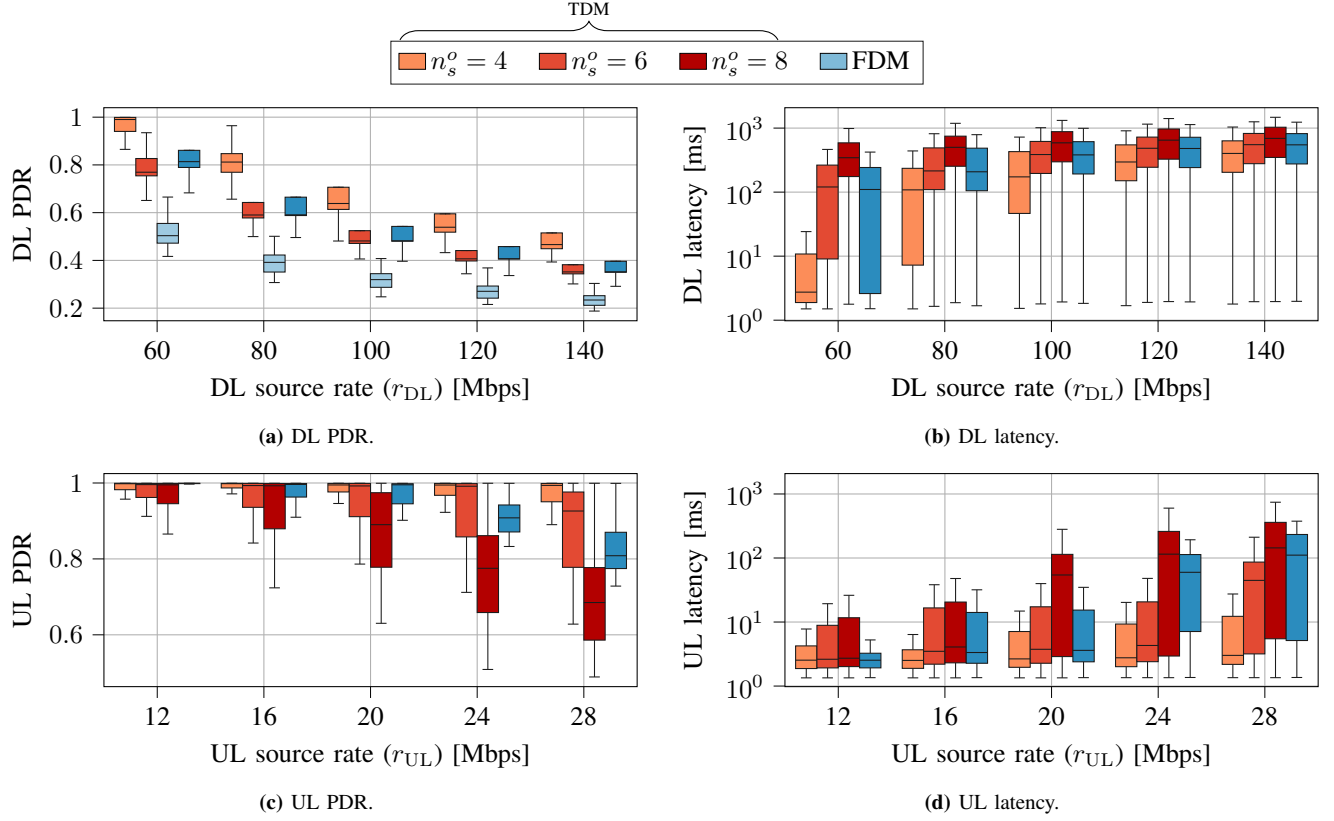

    \begin{subfigure}[b]{\linewidth}
	\centering
	\setlength\fwidth{\columnwidth}
%
%

\definecolor{color1}{RGB}{252,141,89}
\definecolor{color2}{RGB}{227,74,51}
\definecolor{color3}{RGB}{179,0,0}
\definecolor{color4}{RGB}{43,140,190}

\begin{tikzpicture}

\draw [decorate,decoration={brace,mirror,amplitude=7pt},xshift=0pt,yshift=-0.2cm](1.6,0.5) -- (-3.4,0.5) node[black,midway,above,xshift=0cm,yshift=0.2cm] 
{\scriptsize TDM};

\pgfplotsset{every tick label/.append style={font=\scriptsize}}

\pgfplotsset{compat=1.11,
	/pgfplots/ybar legend/.style={
		/pgfplots/legend image code/.code={%
			\draw[##1,/tikz/.cd,yshift=-0.25em]
			(0cm,0cm) rectangle (10pt,0.6em);},
	},
}

\begin{axis}[%
width=0,
height=0,
at={(0,0)},
scale only axis,
xtick={},
ytick={},
axis background/.style={fill=white},
legend style={legend cell align=left,
              align=center,
              draw=white!15!black,
              at={(0.5, 1.3)},
              anchor=center,
              /tikz/every even column/.append style={column sep=1em}},
legend columns=4,
]
\addplot[ybar,ybar legend,draw=black,fill=color1,line width=0.08pt]
table[row sep=crcr]{%
	0	0\\
};
\addlegendentry{$n^{o}_{s}=4$}

\addplot[ybar,ybar legend,draw=black,fill=color2,line width=0.08pt]
table[row sep=crcr]{%
	0	0\\
};
\addlegendentry{$n^{o}_{s}=6$}

\addplot[ybar,ybar legend,draw=black,fill=color3,line width=0.08pt]
table[row sep=crcr]{%
	0	0\\
};
\addlegendentry{$n^{o}_{s}=8$}

\addplot[ybar legend,ybar,draw=black,fill=color4,fill opacity=0.6,line width=0.08pt]
  table[row sep=crcr]{%
	0	0\\
};
\addlegendentry{FDM}

\end{axis}
\end{tikzpicture}%
    \end{subfigure}
    \vskip 0.2cm
    \centering
    \subfloat[][DL PDR.]
	{
	        \label{fig:pdr_dl_mul}
            \input{Figs/PDR_0.2_3_dl_fdm}
	}
    \subfloat[][DL latency.]
	{
            \label{fig:latency_dl_mul}
            \input{Figs/latency_dl_3_log_fdm}
	}
    \vskip 0.2cm
    \subfloat[][UL PDR.]
	{
            \label{fig:pdr_ul_mul}
            \input{Figs/PDR_0.2_3_ul_fdm}
	}
    \subfloat[][UL latency.]
	{
            \label{fig:latency_ul_mul}
            \input{Figs/latency_ul_3_log_fdm}
	}
    \vskip 0.2cm
    \caption{DL and UL PDR and latency as a function of the source rate for different multiplexing schemes in Topology 3.}
   \label{fig:mul}
   \vskip -0.1em
\end{figure*}

\subsection{MT and DU Multiplexing}
\label{sub:multiplexing-eval}
In this section, we compare different \gls{mt} and \gls{du} multiplexing mechanisms in Topology 3, considering a \gls{ul} source rate equal to $r_{\rm UL} = r_{\rm DL}/5$.
We compare \gls{tdm} (for different values of $n^{o}_{s}$) vs. \gls{fdm} with a balanced bandwidth partition between \gls{du} and \gls{mt}.
As described in~\cref{sec:duplexing}, in \gls{tdm}, $n^{o}_{s}$ denotes the number of \gls{ofdm} symbols reserved to odd-layer \gls{iab}-nodes. In Topology 3, this includes layer-1 and layer-3 \gls{iab}-nodes.
The remaining $12 - n^{o}_{s}$ \gls{ofdm} symbols are assigned to even-layer \gls{iab}-nodes, which in Topology 3 corresponds to the \gls{iab}-donor and layer-2 \gls{iab}-nodes.

In Topology 3, backhaul links from the \gls{iab}-donor to layer-1 \gls{iab}-nodes need to carry the whole aggregated traffic generated and requested by all \gls{iab}-nodes in the network.
Therefore, in~\cref{fig:mul} we observe that reducing $n^{o}_{s}$ (thus increasing the number of \gls{ofdm} symbols that the \gls{iab}-donor can allocate) improves the overall network performance in terms of both PDR and latency.
On the contrary, setting $n^{o}_{s}=8$ increases the capacity between layer-1 and layer-2 \gls{iab}-nodes, but reduces the capacity of the \gls{iab}-donor, which becomes the bottleneck.
For example, in \cref{fig:latency_dl_mul,fig:latency_ul_mul}, we focus on the \gls{dl} and \gls{ul} latency as a function of the source rate.
With $r_{\rm DL} = 60$~Mbps, $r_{\rm UL} = 12$~Mbps, and $n^{o}_{s} = 4$, the median \gls{dl} latency is below 10~ms, but it rises sharply to approximately 500~ms when $n^{o}_{s}=8$.
The same conclusions can be drawn for the UL \gls{pdr} and latency, which are consistently close to 1 and lower than 10~ms, respectively, for $n^{o}_{s} = 4$ even when the source rate increases.
These results demonstrate the critical role of resource allocation between \glspl{mt} and \glspl{du} in IAB.
Finally, FDM performs similarly to \gls{tdm} with $n^{o}_{s}=6$, given that the \gls{du} and \gls{mt} bandwidth is partitioned equally.

\section{Conclusions and Future Work}
\label{sec:conclusions}

In this paper, we presented a comprehensive study of \gls{5g} \gls{iab} networks, focusing on their performance in maritime environments.
To this end, we developed and released an open-source, ns-3-based simulator fully aligned with the \gls{3gpp} Release 16 \gls{iab} specifications~\cite{171880}.
The framework integrates a dedicated maritime channel model capturing sea-surface reflections and a rain attenuation model, and supports flexible configuration of the control overhead, slot formats, and multiplexing mechanisms including both \gls{tdm} and \gls{fdm}.

Through an extensive system-level simulation campaign, we observed several key insights.
The performance of maritime \gls{iab} networks is strongly influenced by the deployment topology, weather conditions, and resource allocation strategies. Single-hop deployments offer lower latency and higher \gls{pdr} under clear weather conditions, but are highly susceptible to rain attenuation.
Multi-hop topologies, despite the resulting interference, are characterized by shorter link lengths and therefore experience a higher \gls{snr}. Interestingly, heavy rain can attenuate interfering signals more than the desired ones, which may improve the network performance.
These complementary behaviors indicate that no single topology is globally optimal, suggesting that dynamically reconfiguring the topology based on external factors such as weather and channel conditions could enable the network to maintain or even enhance the overall performance.

Resource allocation also plays a central role.
The configuration of \gls{dl} and \gls{ul} slots, and the distribution of resources between \glspl{du} and \glspl{mt}, significantly affect the network performance.
In multi-hop topologies, prioritizing resource allocation to the IAB-donor is essential to avoid bottleneck effects, as the donor must carry the aggregated traffic from all connected IAB-nodes.
Moreover, our results show that, under equal bandwidth partition, \gls{fdm} achieves similar throughput to \gls{tdm}.
This is expected, as the total amount of resources allocated to each node is the same in both schemes.
However, \gls{fdm} offers greater flexibility by enabling simultaneous parent-child operations, which can reduce latency and simplify scheduling in multi-hop topologies.

The proposed simulator opens several future research directions.
It can be used to validate theoretical models through end-to-end simulations, and support \gls{iab} network optimization.
The simulator can also be adapted to other non-terrestrial environments, such as satellite- or aerial-based networks.


\balance

\bibliographystyle{IEEEtran}
\bibliography{biblio.bib}

\onecolumn

\end{document}